\documentclass[twocolumn,pra,aps,superscriptaddress,showpacs]{revtex4-1}

\usepackage{lineno}
  \usepackage{mathptmx}
\usepackage{subfigure}
\usepackage{dcolumn}
\usepackage{amsmath,amssymb}
\usepackage{bm}
\usepackage{color}
\usepackage{overpic}
\usepackage{latexsym}
\usepackage{epstopdf}
\usepackage{color}
\usepackage[english]{babel}
\usepackage{latexsym}
\usepackage{stmaryrd}

\usepackage{psfrag,graphicx} 
\usepackage{epsf} 
\usepackage{subfigure} 
\usepackage{amsmath} 
\usepackage{amssymb} 
\usepackage{amsfonts}
\usepackage{bm}
\usepackage{natbib}
\usepackage{epstopdf}
\usepackage{appendix}

\definecolor{mygrey}{gray}{0.35}
\definecolor{myblue}{rgb}{0.2,0.2,0.8}
\definecolor{myzard}{cmyk}{0,0,0.05,0}
\definecolor{mywhite}{rgb}{1,1,1}
\definecolor{myred}{rgb}{1,0.,0.3}

\usepackage[colorlinks=true,citecolor=myblue,linkcolor=myred]{hyperref}
\def\be{\begin{equation}}
\def\ee{\end{equation}}
\def\ba{\begin{align}}
\def\enda{\end{align}}
\def\bi{\begin{itemize}}
\def\ei{\end{itemize}}

 \def\ee{\mathord{\rm e}}
 
 \def\ii{\mathord{\rm i}}

\def\half{\textstyle\frac{1}{2}}

 \def\ee{\mathord{\rm e}}
 
 \def\ii{\mathord{\rm i}}

\def\half{\textstyle\frac{1}{2}}

\renewcommand{\ii}{{\rm i}}
\renewcommand{\ee}{{\rm e}}

\def\beq{\begin{equation}}
\def\beq{\begin{equation}}
\def\eeq{\end{equation}}

 \newcommand{\ket}[1]{|#1\rangle}
 \newcommand{\bra}[1]{\langle #1|}

\begin{document}


\title[Short Title]{Spin-Peierls Quantum Phase Transitions in Coulomb Crystals}

\author{A. Bermudez}
\affiliation{Institut f\"{u}r Theoretische Physik, Albert-Einstein Allee 11, Universit\"{a}t Ulm, 89069 Ulm, Germany}
\author{M. B. Plenio}
\affiliation{Institut f\"{u}r Theoretische Physik, Albert-Einstein Allee 11, Universit\"{a}t Ulm, 89069 Ulm, Germany}

\pacs{ 03.67.Ac, 75.10.Jm, 03.67.-a, 37.10.Vz}
\begin{abstract}
The spin-Peierls instability describes a structural transition of a crystal due to strong 
magnetic interactions. Here we demonstrate that cold Coulomb crystals of trapped 
ions provide an experimental testbed in which to study this complex many-body 
problem and to access extreme regimes where the instability is triggered by quantum 
fluctuations alone. We present a consistent analysis based on different analytical and
numerical methods, and provide a detailed discussion of its feasibility on the basis of 
ion-trap experiments. Moreover, we identify regimes where this quantum simulation 
may exceed the power of classical computers.

\end{abstract}

\maketitle

The beauty of low-dimensional quantum many-body systems (QMBS) lies in the complexity born of the combination of interactions, disorder, and quantum fluctuations. However, these ingredients also conspire to render perturbative techniques inefficient, posing thus a fundamental challenge that has inspired the development of a variety of analytical~\cite{tsvelik_book} and numerical~\cite{numerics} tools. Moreover, the synthesis of low-dimensional materials has upgraded these challenges from a theoretical endeavor into a discipline that underlies some of the most exciting recent discoveries in condensed-matter physics, such as the fractional quantum Hall effect. The recent progress in the field of atomic, molecular, and optical (AMO) physics presents a promising alternative to these solid-state realizations of low-dimensional QMBS. This field, which was originally devoted to the study of light-matter interactions at the scale of a single or few atoms, is progressively focusing on the many-body regime in  platforms such as neutral atoms in optical lattices~\cite{review_qs_atoms}, cold  Coulomb crystals of trapped ions~\cite{review_qs_ions}, or coupled cavity arrays~\cite{coupled_cavities}. The possibility of experimentally  designing  the microscopic  Hamiltonians  in order to target a variety of complicated many-body models introduces a novel approach to explore QMBS in a controlled fashion, the so-called {\it quantum simulations} (QSs)~\cite{qs}. Some  remarkable QSs in AMO platforms are the optical-lattice realization of Mott~\cite{mott} and Anderson~\cite{anderson} insulators,  and the recent efforts towards a quantum-Hall insulator~\cite{hall}. More specific to the subject of this manuscript is the QS of  quantum magnetism in trapped ions~\cite{ising_exp}, and  more recently in optical lattices~\cite{ising}.

In this Letter, we explore the capabilities of AMO setups for the QS of interaction-mediated instabilities in QMBS. The standard playground for these phenomena  is the so-called metal-insulator transition~\cite{metal_insulator}, which has been investigated for a variety of transition-metal compounds in the field of strongly-correlated electrons. A paradigmatic case is the one-dimensional  metal, where either the electron-electron interactions destabilize the metal towards a superconducting state, or  the electron-phonon coupling leads to a charge-density-wave condensate~\cite{cdw_book}. The latter instability is a consequence of the so-called {\it Peierls transition}~\cite{peierls_transition}, where the electron-phonon interactions induce a periodic distortion of the ionic lattice, and open an energy gap in the conduction band  of the metal. By virtue of the Jordan-Wigner transformation~\cite{jw}, this phenomenon finds a magnetic counterpart: the {\it spin-Peierls transition}~\cite{spin_peierls_heisenberg}, whereby a spin-phonon-coupled antiferromagnet becomes unstable with respect to a dimerization of the lattice. This creates an alternating pattern of weak and strong spin interactions,  which in turn opens an energy gap in the spectrum of  collective excitations.  We note that this instability has turned out to be important for different compounds, such as organic molecular crystals and transition-metal oxides~\cite{supp_material_perspective}.

\begin{figure}

\centering
\includegraphics[width=0.9\columnwidth]{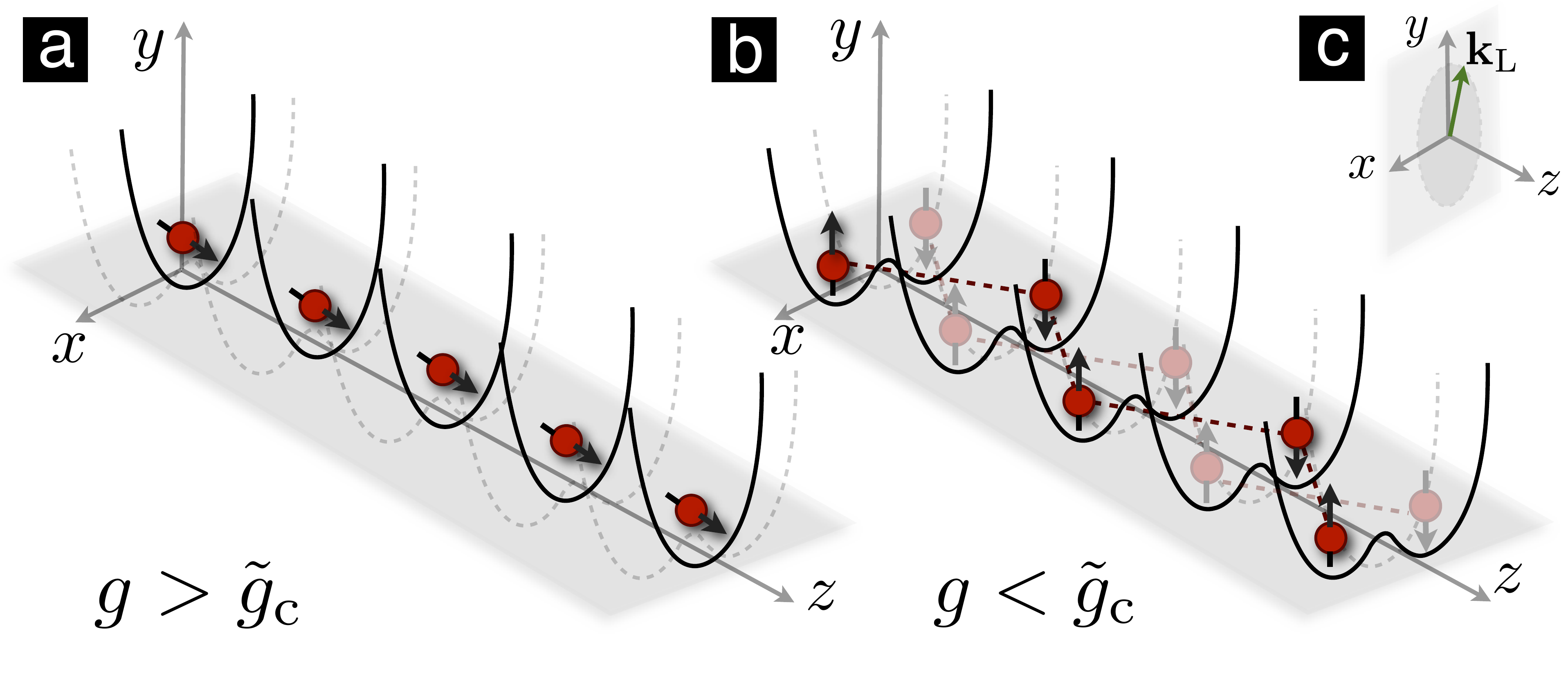}
\caption{ {\bf Scheme of the spin-Peierls instability:}   {\bf (a)} In the  paramagnetic phase  $\ket{{\rm P}}=\ket{\uparrow\uparrow\cdots\uparrow}$ all the spins are parallel to a transverse field $g>\tilde{g}_{\rm c}$ that points along the $z$-axis, and the Coulomb crystals corresponds to an ion string. {\bf (b)} The antiferromagnetic phase  $g<\tilde{g}_{\rm c}$ corresponds to the two  N\'eel-ordered groundstates $\ket{{\rm AF}}\in\{\ket{+-\cdots+-},\ket {-+\cdots-+}\}$, where the spins are antiparallel in the $x$-basis  $\ket{\pm}=(\ket{\!\uparrow}\pm\ket{\!\downarrow})/\sqrt{2}$.  This order-disorder quantum phase transition occurs with the linear-to-zigzag structural phase transition. {\bf (c)} Arrangement of the laser wavevector ${\bf k}_{\rm L}$ lying within the $xy$-plane. }
\label{spin_peierls_scheme}
\end{figure}

From a theoretical perspective, the complete understanding of such a complex many-body  system, treating the dynamics of the spins and phonons on the same footing, is still considered to be an open problem~\cite{supp_material_perspective,giamarchi_book}. From an experimental point of view, the  spin-Peierls instabilities observed  so far~\cite{supp_material_perspective,spin_peierls_heisenberg_exp} take place at finite temperatures, and are limited to the so-called Heisenberg model. Hence, the possibility of  realizing a spin-Peierls transition only driven  by  quantum fluctuations, and possibly exploring different types of magnetic interactions, remains as an experimental challenge. 

We hereby present a theoretical proposal for a trapped-ion QS to tackle both problems. In particular, by building on the recent experiments~\cite{ising_exp} on the quantum Ising model (QIM)~\cite{tfi}, we describe how to tailor a spin-Peierls instability. We show that {\it (i)} the disordered paramagnet in a linear ion chain changes into an ordered antiferromagnet in a zigzag  crystal [Fig.~\ref{spin_peierls_scheme}{\bf (a)-(b)}], and {\it (ii)} the spin-Peierls transition can be driven only by the quantum fluctuations introduced by the transverse field of the QIM. Let us remark that, in comparison to neutral atoms in optical lattices or coupled arrays of  cavities, trapped ions seem to be the best candidates to realize the spin-Peierls quantum simulator. One of the main reasons is that the underlying lattice is not  externally fixed, but rather results from the self-assembling dynamics of the ions.

{\it The system.--} The advent of  experimental techniques for the confinement, cooling, and coherent manipulation of  atomic ions has recently been exploited for QS purposes~\cite{review_qs_ions},  where the controlled increase of the number of trapped ions yields a genuine bottom-up approach to the many-body regime. We consider a Coulomb gas formed by an ensemble of $N$ trapped ions of mass $m$, and charge $e$,  which are described  by the Hamiltonian
 \begin{equation}
 \label{coulomb_gas}
 H_0=\frac{\omega_0}{2}\sum_i\sigma_i^z+\sum_{i,\alpha}\left(\frac{1}{2m}p_{i\alpha}^2+\frac{1}{2} m\omega_{\alpha}^2r_{i\alpha}^2\right)+\frac{e^2}{2}\sum_{i\neq j}\frac{1}{|{\bf r}_i-{\bf r}_j|}
 \end{equation}
where $\{\omega_{\alpha}\}_{\alpha=x,y,z}$ are the effective trapping frequencies of a linear Paul trap.  Here, $\omega_0$ is the energy difference between two electronic groundstates of the atomic structure $\ket{\!\uparrow}_i,\ket{\!\downarrow}_i$, where $\hbar=1$, and $\sigma_i^z=\ket{\!\uparrow_i}\bra{\uparrow_i\!\!}-\ket{\!\downarrow_i}\bra{\downarrow_i\!\!}$. This Hamiltonian must be complemented by the laser-ion interaction responsible for  coupling the electronic and the motional degrees of freedom. We consider a pair of laser beams with frequencies $\{\omega_l\}_{l=1,2}$,  wavevectors $\{{\bf k}_l\}_{l=1,2}$, and phases $\{\phi_l\}_{l=1,2}$, which are tuned close to the atomic transition.  In the dipolar approximation,   the laser-ion Hamiltonian becomes
   \begin{equation}
   \label{laser_ion}
   H_{\rm L}=\sum_{l,i}(\Omega_l\sigma_i^++\Omega_l^*\sigma_i^-)\cos({\bf k}_l\cdot{\bf r}_i-\omega_lt+\phi_l),
   \end{equation}
where $\Omega_l$ stands for the Rabi frequency of the transition, and we have introduced the spin raising and lowering operators $\sigma_i^+=\ket{\!\uparrow_i}\bra{\downarrow_i\!}=(\sigma_i^-)^{\dagger}$. To proceed further, we need to make some assumptions about the dynamics of this atomic plasma.

   As evidenced in early experiments~\cite{structural_phase_transition}, a laser-cooled  ensemble of ions self-assembles in a Coulomb crystal, which undergoes a series of structural phase transitions (SPTs) as the trapping conditions are modified. In particular, when $\omega_y\gg\omega_{x},\omega_z,$ and the ratio $\kappa_x=(\omega_z/\omega_x)^2$ is tuned across a critical value $\kappa_{\rm c}$~\cite{structural_transition}, the geometry of the crystal  changes from a linear string to a zigzag ladder. Note that this SPT   displays a rich phenomenology that has recently revived the interest in the subject~\cite{landau_theory,zig_zag_theory,zigzag_inhomogeneous,rg_zigzag,ising_theory,jahn_teller}. Here, we focus on the linear regime close to the critical point $\kappa_x\lesssim\kappa_{\rm c}$, where the vibrations of the ions along each of the confining axes are decoupled. We consider that the laser wavevectors in~\eqref{laser_ion} lie within the $xy$-plane [Fig.~\ref{spin_peierls_scheme}{\bf (c)}], ${\bf k}_l=k_{lx}{\bf e}_x+k_{ly}{\bf e}_y$, such that their  frequencies are tuned close to the resonance of the vibrational sidebands $\omega_l\approx\omega_0\pm\omega_y$. Since these correspond to the strongly-confining $y$-axis, $\omega_y\gg\omega_x,\omega_z$, the coupling of the laser beams to the $x$ and $z$ vibrational modes becomes far off-resonant and can  be neglected. Let us remark that this argument has one possible exception, there might be a vibrational soft mode where the phonons condense at the SPT. As identified in~\cite{landau_theory}, this corresponds precisely to the zigzag mode along the $x$-axis.   This affects  the laser-ion coupling~\eqref{laser_ion} regardless of the soft-mode frequency (see Supplementary Material).
   
   The above considerations allow us  to extract the relevant part of the Hamiltonian~\eqref{coulomb_gas} after introducing  ${\bf r}_i=l_z(\tilde{z}_i^0{\bf e}_z+ q_{ix}{\bf e}_x+q_{iy}{\bf e}_y+q_{iz}{\bf e}_z)$, where $\tilde{z}_i^0$ are the equilibrium positions  in units of $l_z=(e^2/m\omega_z^2)^{1/3}$, and $q_{i\alpha}$ are the small displacements along the corresponding axes. Following~\cite{ising_theory}, the displacements along the $x$-axis have  been adapted to the aforementioned zigzag mode $q_{ix}=(-1)^i\delta q_{ix}$, such that $\delta q_{ix}$ is a smooth function that allows a gradient expansion. The Hamiltonian then becomes  $H_0=\frac{1}{2}\sum_i\omega_0\sigma_i^z+H_{x}+H_{y}$, where 
 \begin{equation}
 \label{oscillators}
 \begin{split}
 H_{x}&=\!\sum_i\!\!\left(\!\frac{ml_z^2}{2}\!\left(\partial_t \delta q_{ix}\right)^2\!+\frac{r^{x}_i}{2}\delta q_{ix}^2\!+\frac{u^{x}_i}{4}\delta q_{ix}^4\!\right)\!+\!\sum_{i\neq j}\!\frac{K^{x}_{ij}}{2}(\partial_j\delta q_{ix})^2\!,\\
H_{y}&=\sum_i\left(\frac{ml_z^2}{2}\left(\partial_t q_{iy}\right)^2+\frac{r^{y}_i}{2} q_{iy}^2\right)+\sum_{i\neq j}\frac{K^{y}_{ij}}{2}(\partial_jq_{iy})^2,
\end{split}
 \end{equation}
and we have introduced the gradient $\partial_j f_i= f_i- f_j$. In these expressions, the  coupling energies for the vibrations are
  \begin{equation}
 \label{couplings_x}
 \begin{split}
 & r^x_{i}=m\omega_x^2l_z^2\left(1-\half\kappa_x\zeta_i(3)\right),\hspace{1ex} u_i^x=m\omega_x^2l_z^2\left(\frac{3}{4}\kappa_x\zeta_i(5)\right),
\\
 & K_{ij}^x=m\omega_x^2l_z^2\left(\sum_{l\neq i}\frac{(-1)^{l+i+1}\kappa_x}{2|\tilde{z}_i^0-\tilde{z}_l^0|}\right)\delta_{j,i+1},\\
 \end{split}
 \end{equation}
along the $x$-axis, expressed in terms of the inhomogeneous function  $\zeta_i(n)=\sum_{l\neq i}[(-1)^i-(-1)^l]^{n-1}|\tilde{z}_i^0-\tilde{z}_l^0|^{-n}$ with $n\in\mathbb{Z}$, and the Kronecker delta $\delta_{lm}$. In the limit of tight confinement along the $y$-axis, $\kappa_y=(\omega_z/\omega_y)^2\ll1$, we find
 \begin{equation}
 \label{couplings_y}
 \begin{split}
 r^{y}_i&=m\omega_y^2l_z^2, \hspace{1.5ex} 
 K^{y}_{ij}=\frac{m\omega_y^2l_z^2\kappa_y}{2|{\tilde{z}}_i^0-{\tilde{z}}_j^0|^3},\\
 \end{split}
 \end{equation}
Accordingly, $H_y$ corresponds to a set of dipolarly-coupled harmonic oscillators, whereas   $H_x$ describes a set of nearest-neighbor-coupled anharmonic oscillators. 

 The coupled harmonic oscillators~\eqref{oscillators}  are  diagonalized yielding a set of collective modes with frequencies $\omega_n$, whose excitations, created and annihilated by $a_n^{\dagger},a_n^{\phantom{\dagger}}$, shall be referred to as the {\it hard phonons}. This yields the quadratic Hamiltonian $H_{y}=\sum_n\omega_na_n^{\dagger}a_n$. By setting the laser frequencies to the red and blue vibrational sidebands of the atomic transition,  $\omega_1\approx\omega_0-\omega_n$ and  $\omega_2\approx\omega_0+\omega_n$, we can express the laser-ion interaction~\eqref{laser_ion} as follows (see Supplementary Material) 
  \begin{equation}
  \label{sidebands}
   H_{\rm L}=\sum_{in}\left(\mathcal{F}_{in}^{\rm r}\ee^{\ii \theta_{\rm r}q_{ix}}\sigma_i^+a_n^{\phantom{\dagger}}+\mathcal{F}_{in}^{\rm b}\ee^{\ii \theta_{\rm b}q_{ix}}\sigma_i^+a_n^{\dagger}+\text{H.c.}\right),
   \end{equation}
   where we have introduced the sideband coupling strengths $\mathcal{F}_{in}^{\rm r}=\frac{\ii}{2}\Omega_1\eta_{1n}\mathcal{M}_{in}\ee^{\ii\phi_1},\mathcal{F}_{in}^{\rm b}=\frac{\ii}{2}\Omega_2\eta_{2n}\mathcal{M}_{in}\ee^{\ii\phi_2}$,  the collective Lamb-Dicke parameters $\eta_{ln}=k_{ly}/\sqrt{2m\omega_n}\ll 1$,  the normal-mode vibrational amplitudes $\mathcal{M}_{in}$, and  $\theta_{\rm r}=k_{1x}l_z,\theta_{\rm b}=k_{2x}l_z$. 
   
  In contrast, the anharmonic oscillators~\eqref{oscillators} correspond to an inhomogeneous version of the $\phi^4$ model on a lattice, namely, an interacting    scalar field theory that cannot be exactly  diagonalized. This model yields a SPT that can be  understood as follows. The regime $r_i^x>0$, $u_i^x>0$ corresponding to  trapping-frequency ratios fulfilling $\kappa_x<\kappa_{{\rm c},i}=2/\zeta_i(3)$, yields the linear ion configuration, which respects the $\mathbb{Z}_2$ symmetry of the model.  Conversely, when $r_i^x<0$, $u_i^x>0$ for $\kappa_x>\kappa_{{\rm c},i}$, the ions self-organize in the the zigzag ladder corresponding to the  broken-symmetry phase, whereby $\langle	\delta q_{ix}\rangle\neq 0$ signals the condensation of the {\it soft phonons} in the zigzag mode. We note that  the $\phi^4$ model fulfills $r_i^x\neq r_j^x$, which leads to an inhomogeneous SPT setting at the center of the trap~\cite{zigzag_inhomogeneous}. As outlined previously,  when the soft phonons condense $\langle\delta q_{ix}\rangle\neq0$, there is a non-trivial effect in the laser-ion Hamiltonian~\eqref{sidebands} that must be considered. We show below that, in this case, the hard phonons mediate a spin-spin interaction, whereas the soft condensed phonons are responsible for a dimerization of the coupling strengths. This turns out to be the key ingredient for a zero-temperature spin-Peierls transition. 
Let us also emphasize that this model, which is a cornerstone in the microscopic description of SPTs~\cite{spt}, has not been combined with a spin-Peierls distortion to the best of our knowledge. 
   
  {\it Dimerized quantum spin model.--} The spin-phonon  model in~\eqref{oscillators} and~\eqref{sidebands} yields an extremely complex QMBS. We  analyze the onset of a spin-Peierls quantum phase transition by performing a series of simplifications. First, we  neglect the time-dependence of the zigzag distortion~\eqref{oscillators}. This adiabatic approximation, which is standard in the treatment of spin-Peierls phenomena~\cite{spin_peierls_heisenberg}, is justified if the zigzag mode is much slower than the effective spin dynamics, which is valid close to the critical point.  Hence, we treat the SPT classically by setting $q_{ix}= (-1)^i\langle \delta q_{ix}\rangle$ self-consistently. Second, we consider a homogeneous zigzag distortion, which amounts to neglecting the nearest-neighbor couplings in Eq.~\eqref{oscillators}.    
Third, when the coupling of the spins to the hard phonons~\eqref{sidebands} is weak, they can be integrated out yielding an effective quantum spin model. In the linear string, this leads to a dipolar version of the celebrated QIM~\cite{ising_porras}, whereas the frustrated $J_1$-$J_2$ QIM arises in the zigzag configuration~\cite{ising_frustration}.  In this work, we show that in the vicinity of the critical point $\kappa_x\approx\kappa_{{\rm c,}i}$, the quantum spin model corresponds to a dipolar QIM with additional spin-spin couplings whose sign alternates  periodically when the soft phonons condense.

In analogy to the S$\o$rensen-M$\o$lmer gates~\cite{molmer_sorensen_gate}, we consider that the red- and blue-sideband terms~\eqref{sidebands} have opposite detunings  $\delta_{n{\rm r}}=-\delta_{n{\rm b}}=:\delta_n$, where $\delta_{n{\rm r}}=\omega_1-(\omega_0-\omega_n)$, and $\delta_{n{\rm b}}=\omega_2-(\omega_0+\omega_n)$. Besides, their Rabi frequencies fulfill $\Omega_1k_{1y}^2=\Omega_2k_{2y}^2$, and attain values such that $|\mathcal{F}_{in}^{\rm r}|=|\mathcal{F}_{in}^{\rm b}|\ll\delta_n$. In this limit, the  sidebands~\eqref{sidebands}  create a virtual hard phonon which is then reabsorbed by a distant ion,  leading thus to the aforementioned effective spin-spin interaction. The above constraints  are responsible for the destructive interference of the processes where a phonon is created and then reabsorbed by the same ion, a crucial property that underlies the availability of an effective spin Hamiltonian that is decoupled from the hard phonons. Finally, by considering that the pair of  laser beams are counter-propagating  ${\bf k}_1=-{\bf k}_2=:{\bf k}$, which implies that the parameters are  $\theta_{\rm r}=-\theta_{\rm b}=:\theta$, it is possible to obtain the following spin Hamiltonian (see Supplementary Material)
\begin{equation}
\label{xy_gen}
H_{\rm eff}=\sum_{i\neq j}\left(J_{ij}^{xx}\sigma_i^x\sigma_j^x+J_{ij}^{yy}\sigma_i^y\sigma_j^y+J_{ij}^{xy}\sigma_i^x\sigma_j^y+J_{ij}^{yx}\sigma_i^y\sigma_j^x\right),
\end{equation}
where the coupling strengths are the following
\begin{equation}
\label{couplings}
\begin{split}
J_{ij}^{xx}&=J_{ij}\big(\cos(\theta(q_{ix}-q_{jx}))+\cos\left(\theta(q_{ix}+q_{jx})+\phi_-\right)\big),\\
J_{ij}^{yy}&=J_{ij}\big(\cos(\theta(q_{ix}-q_{jx}))-\cos\left(\theta(q_{ix}+q_{jx})+\phi_-\right)\big),\\
J_{ij}^{xy}&=J_{ij}\big(\sin(\theta(q_{ix}-q_{jx}))-\sin(\theta(q_{ix}+q_{jx})+\phi_-)\big),\\
J_{ij}^{yx}&=-J_{ij}\big(\sin(\theta(q_{ix}-q_{jx}))+\sin(\theta(q_{ix}+q_{jx})+\phi_-)\big),\\
\end{split}
\end{equation}
and we have introduced the relative phase between the lasers $\phi_-=\phi_{1}-\phi_2$. Here, the spin-spin coupling strengths are 
\begin{equation}
\label{dipolar_coupling}
J_{ij}=\frac{J_{\rm eff}}{2|\tilde{z}_i^0-\tilde{z}_j^0|^3},\hspace{1.5ex}J_{\rm eff}=\frac{\Omega_{\rm L}^2\eta_y^2\kappa_y}{16\delta_y^2}\left(1+\frac{\delta_y}{\omega_y}\right)\omega_y,
\end{equation}
where we made a expansion for $\kappa_y\ll1$, and introduced the bare detuning  $\delta_y=\omega_1-(\omega_0-\omega_y)=-\omega_2+(\omega_0+\omega_y)$, the bare Lamb-Dicke parameter $\eta_y=k_y/\sqrt{2m\omega_y}$, and the common Rabi frequency $\Omega_{\rm L}:=\Omega_1=\Omega_2$. Hence, the spin-phonon interaction leads to a generalization of the famous XY quantum spin model~\cite{xy_model}. In order to obtain the promised dimerized spin model, one has to  phase-lock the laser beams $\phi_-=0$, and consider  the critical region where the zigzag distortion is small enough $\theta\langle\delta q_i\rangle\ll 1$. Then, we can Taylor expand and obtain an  antiferromagnetic Ising interaction characterized by $J_{ij}^{xx}=2J_{ij}$, and $J_{ij}^{yy}=0$, but also
\begin{equation}
\label{coupling_strengths}
J_{ij}^{xy}=2J_{ij}(-1)^{j+1} \theta\langle\delta q_{jx}\rangle=J_{ji}^{yx},
\end{equation}
which give rise to the quantum dimerization (i.e. a magnetic interaction that does not commute with the Ising coupling, and alternates between a ferromagnetic-antiferromagnetic sign $J_{ij}^{xy},J_{ji}^{yx}\propto (-1)^{i+1}$). Besides, we also consider a transverse field $h$ that can be obtained from a microwave that is far off-resonant with respect to the atomic transition. Altogether, the spin Hamiltonian becomes $H_{\rm eff}=H_{\rm Ising}+H_{\rm dimer}$, where
\begin{equation}
\label{dimerized_qim}
\begin{split}
H_{\rm Ising}&=\sum_{i\neq j}\frac{J_{\rm eff}}{|\tilde{z}_i^0-\tilde{z}_j^0|^3}\sigma_i^x\sigma_j^x-h\sum_i\sigma_i^z,\\
H_{\rm dimer}&=\sum_{i\neq j}\frac{J_{\rm eff}(-1)^{j+1}\xi_j}{|\tilde{z}_i^0-\tilde{z}_j^0|^3}\sigma_i^x\sigma_j^y+\frac{J_{\rm eff}(-1)^{i+1}\xi_i}{|\tilde{z}_i^0-\tilde{z}_j^0|^3}\sigma_i^y\sigma_j^x,
\end{split}
\end{equation}
and we have introduced $\xi_i=\theta\langle \delta q_i\rangle\ll 1$. Let us remark that the couplings of the dimerization Hamiltonian depend on the condensation of the soft phonons $\langle\delta q_{ix}\rangle$, which is in turn described by the  $\phi^4$  theory. Below, we show how this model leads to the desired Spin-Peierls instability.

\begin{figure}
\centering
\includegraphics[width=1\columnwidth]{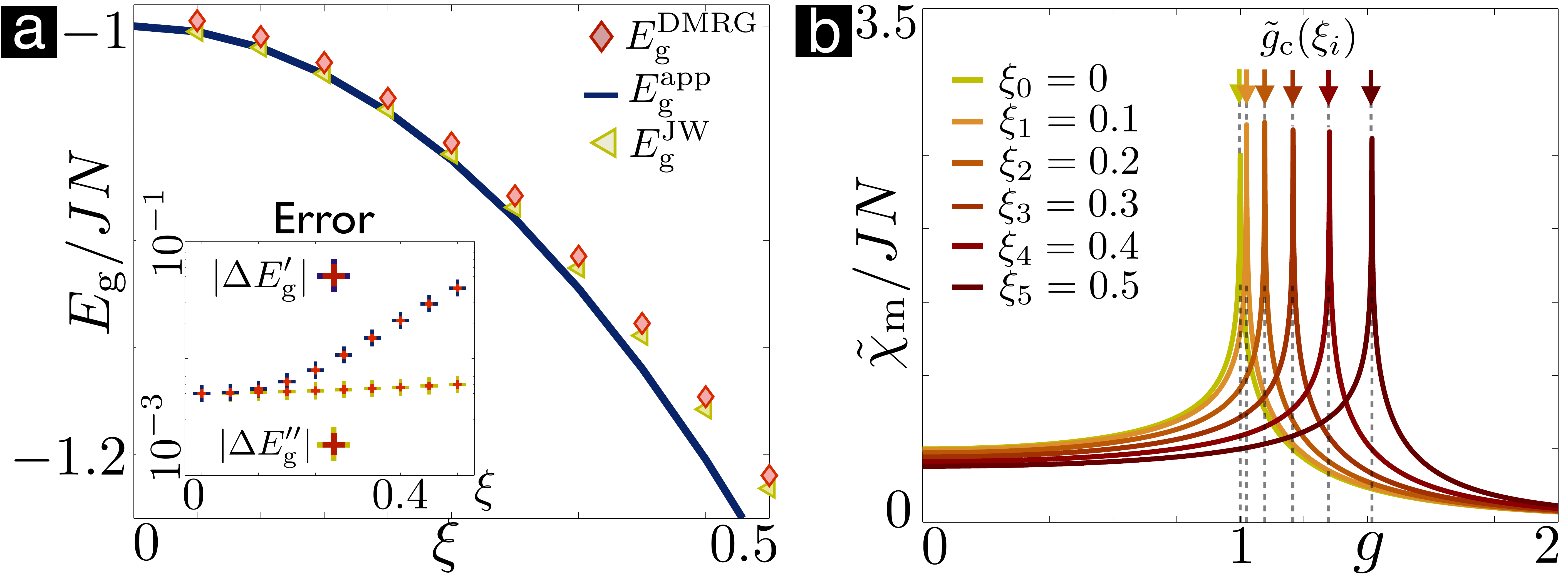}
\caption{ {\bf Spin-Peierls transition:}  {\bf (a)} Scaling of the ground-state energy with the lattice dimerization. {\bf (b)} Displacement of the critical point calculated from the divergence of the magnetic susceptibility. }
\label{spin_peierls}
\end{figure}

{\it Spin-Peierls quantum phase transition.--}  To demonstrate that the introduced scheme yields a QS of the spin-Peierls instability, we  simplify the model by neglecting its long-range interactions and inhomogeneities. The  spin model can be  solved by means of a Jordan-Wigner transformation after setting  $\xi=\xi_i\hspace{1ex}\forall i$, which is later used as input to the $\phi^4$ model self-consistently. In the limit $\xi\ll1$, the groundstate energy fulfills
 \begin{equation}
 \label{energy_decrease}
 E_{\rm g}(\xi)\approx E_{\rm g}(0)-\frac{2JN}{\pi}\varphi(g)\xi^2<E_{\rm g}(0),
\end{equation}
where $J>0$ is the nearest-neighbor antiferromagnetic coupling,  $g=h/J$, and  we have introduced a monotonically-decreasing positive-definite function $\varphi(g)$ that depends on the complete elliptic integrals (see Supplementary Material). We have compared this expression to numerical DMRG~\cite{dmrg_white} calculations (see Fig.~\ref{spin_peierls}{\bf (a)} and Supplementary Material), which  support our claim for  small dimerizations. The above lowering of the groundstate energy pinpoints the instability towards the lattice distortion. Besides,  the spectrum of magnetic excitations  displays the following energy gap $\Delta\propto\big|g-\sqrt{1+4\xi^2}\big|$. With respect to the paramagnetic-to-antiferromagnetic quantum phase transition of the standard QIM at $g_{\rm c}=1$~\cite{tfi}, the dimerization breaks the self-duality of the model and allows the critical point  to flow towards    higher values 
\begin{equation}
\label{qim_cp}
g_{\rm c}\to \tilde{g}_{\rm c}(\xi)=\sqrt{1+4\xi^2}.
\end{equation}
 In Fig.~\ref{spin_peierls}{\bf (b)}, we obtain the critical point from the divergence of the magnetic susceptibility $\chi_{\rm m}\propto\tilde{\chi}_{\rm m}=-\partial^2 E_g/\partial g^2$, which clearly shows that the critical point $g_{\rm c}=1$ flows towards higher values $\tilde{g}_{\rm c}(\xi)$ as the dimerization increases.

Therefore, the disordered paramagnet  close to criticality $g_{\rm c}<g\leq\tilde{g}_{\rm c}$ will be unstable towards the ordered antiferromagnet if the lowering of the energy~\eqref{energy_decrease} compensates the structural change. For self-consistency, we incorporate this energy change in the $\phi^4$ model. By noticing that it  fulfills  $
 E_{\rm g}(\xi)-E_{\rm g}(0)\propto\sum_i\delta q_{ix}^2$, it becomes clear how to modify the parameters of the scalar field theory~\eqref{couplings_x}. In analogy to the magnetic  quantum phase transition, the critical value of the SPT is also displaced,  but this time  towards a smaller value
 \begin{equation}
 \label{spt_cp}
 \kappa_{{\rm c},i}\to\tilde{\kappa}_{{\rm c},i}=\left(\frac{\zeta_i(3)}{2}+\frac{2J}{m\omega_z^2l_z^2}\frac{\theta^2\varphi(g)}{\pi}\right)^{-1}.
 \end{equation}
Therefore, the linear ion string close to criticality $\tilde{\kappa}_{{\rm c},i}\leq \kappa_x<\kappa_{{\rm c},i}$ is unstable towards the symmetry-broken zigzag phase. 

 We have thus proved  our claim {\it (i)} that the paramagnetic phase in the linear ion string will be unstable towards the antiferromagnetic zigzag ladder [Fig.~\ref{spin_peierls_scheme}]. Moreover, by fixing the ratio of the trapping frequencies in the linear regime, $\kappa_{x}<\kappa_{{\rm c},i}$,   we can drive both the structural and the magnetic phase transitions by only modifying the transverse magnetic field $g$ across $\tilde{g}_{\rm c}$. The necessary condition is that the trapping frequencies are tuned according to the following expression 
 \begin{equation}
 \label{condition}
m\omega_z^2l_z^2=\frac{2J\theta^2\varphi(\tilde{g}_{\rm c})}{\pi \left(\kappa_x^{-1}-\half\zeta_i(3)\right)}.
 \end{equation}
Hence, the zigzag antiferromagnet $g<\tilde{g}_{\rm c}$  is driven onto a linear paramagnet by increasing the quantum fluctuations $g>\tilde{g}_{\rm c}$ (and vice versa). This supports our  claim {\it (ii)} that the spin-Peierls transition can be  driven  by quantum fluctuations alone.

{\it Experimental considerations.--} We focus on $^{25}{\rm Mg}^+$ and select two hyperfine levels for the spin states $\ket{\!\uparrow_i}=\ket{F=2,m_{F}=2},\ket{\!\downarrow_i}=\ket{3,3}$, such that the resonance frequency in~\eqref{coulomb_gas} is $\omega_0/2\pi=1.8$ GHz. We consider a $N=30$ ion register with trapping frequencies $\omega_z/2\pi\approx300$ kHz, $\omega_x/2\pi\approx7$ MHz, and $\omega_y/2\pi=10$ MHz. The phase-locked laser beams leading to~\eqref{laser_ion} are blue-detuned $\delta_y/2\pi\approx$1 MHz, such that the two-photon Rabi frequencies are $\Omega_{\rm L}/2\pi\approx 1$ MHz, and the Lamb-Dicke parameter $\eta_{y}\approx 0.2$. With these values, the required constraints detailed in the Supplementary Material are fulfilled, and we obtain a nearest-neighbor spin coupling with the typical strength  $J=2J_{\rm eff}/|\tilde{z}_i^0-\tilde{z}_{i+1}^0|^3\approx 1\text{ kHz}$ observed in experiments~\cite{ising_exp}. By considering these parameters, the condition~\eqref{condition} imposes the following constraint over the anisotropy $(\kappa_{{\rm c,}i}-\kappa_x)/\kappa_{{\rm c},i}\sim10^{-4}\eta_y^2$, which requires to be sufficiently close to the structural phase transition.
 In practice, the soft radial trapping frequency must be controlled with an accuracy of $\Delta\omega_x\sim10^{-6}\omega_x\approx$1-10 Hz, which coincides with the the precision required to observe quantum effects in the SPT~\cite{ising_theory}.  Provided that this precision is achieved in the experiments, one could optically pump the linear ion register to $\ket{\psi(0)}=\otimes\ket{\uparrow_i}$, and then study its adiabatic evolution towards the AF phase as the transverse field $g(t)$ is decreased. The corresponding AF order can be measured by fluorescence techniques, whereas the structural phase transition could be directly observed in a CCD camera, or inferred from spectroscopy of the vibrational modes. A simpler experiment would require to set the  the anisotropy parameter within the instability regime given by the displacements in Eqs.~\eqref{spt_cp} and \eqref{qim_cp}, which leads to $(\kappa_{{\rm c,}i}-\kappa_x)/\kappa_{{\rm c},i}\sim\eta_y^2\approx 10^{-2}$ and $h/J\approx 10^{-2}$. The linear paramagnet would be directly unstable towards to zigzag AF without the need of adiabatically tuning the transverse field $g(t)$, and the demanding accuracy over the trap frequencies.
  
     {\it Conclusions and Outlook.--}  A sensible QS must address  questions that are difficult to assess by other analytical or numerical methods. In this work, we have proposed a trapped-ion  QS that  fulfills this requirement. In particular, in the regime where non-adiabatic effects of the zigzag distortion become relevant, the complexity of the many-body model in Eqs.~\eqref{oscillators} and~\eqref{sidebands} compromises the efficiency of existing numerical methods. Besides, this QS may address the effects of the inhomogeneities, the long-range dipolar tail of the spin-spin interactions, and the dynamics across such a magnetic structural quantum phase transition. We emphasize that the incorporation of all these effects  make our QS of the utmost interest, which may also find an application in the context of other Wigner crystals~\cite{wigner_crystal}, such as electrons in quantum wires or liquid helium.

{\it Acknowledgements.--} This work was partially supported by the EU STREP projects HIP, PICC, and by the Alexander von Humboldt Foundation. We acknowledge useful discussions with J. Almeida and S. Montangero.


\vspace*{50ex}
\newpage
\section*{Supplementary material}

\appendix

\subsection{Phonon condensation in the linear-to-zigzag Coulomb crystal}
 \label{spt}
 In this Appendix, we present a detailed discussion of the vibrational modes of an inhomogeneous Coulomb crystal close to the linear-to-zigzag SPT. The  description  for the homogeneous case, which applies to  ring  traps or to the center of a linear trap in the thermodynamic limit,  can be found in~\cite{landau_theory}. In this Appendix, we focus on the inhomogeneous crystal.
  
\begin{figure}

\centering
\includegraphics[width=1\columnwidth]{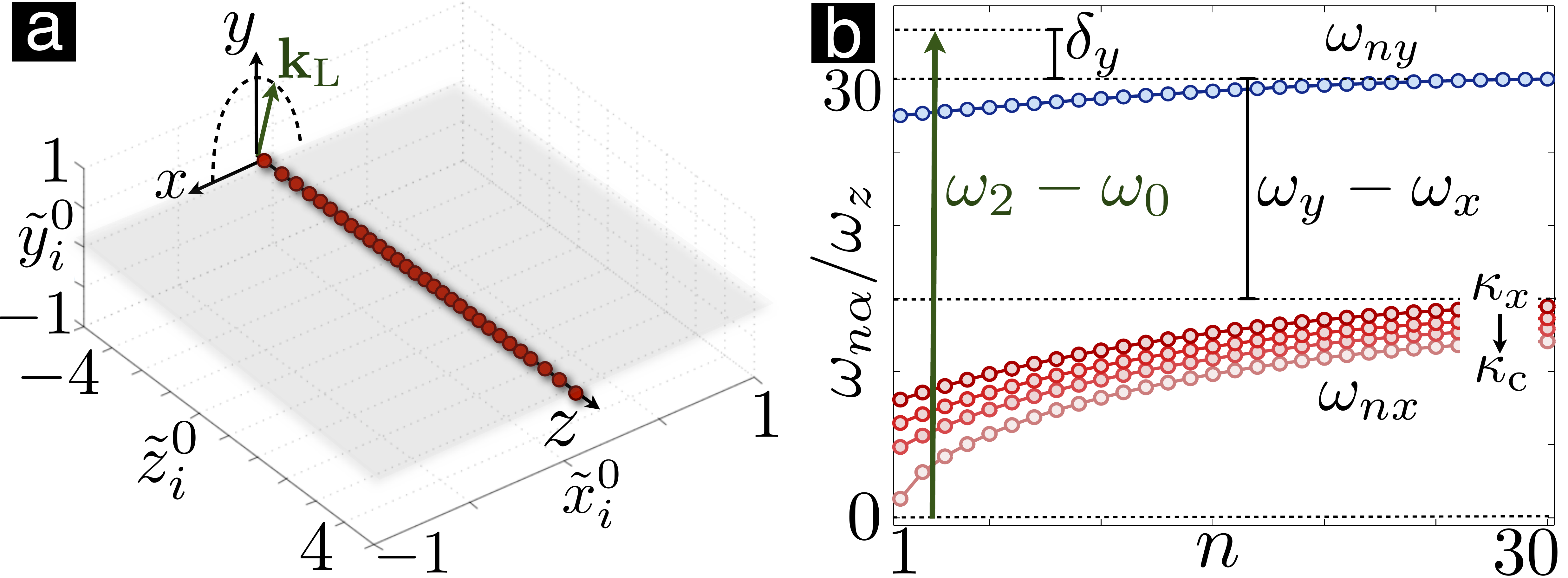}
\caption{ {\bf Radial phonon modes  for a linear ion chain:} {\bf (a)}   Equilibrium positions of linear Coulomb crystal with $N=30$ ions and $\omega_y=30\omega_z$, $\kappa_x\leq\kappa_{\rm c}$. The ions self-assemble in an inhomogeneous string along the $z$-axis, such that the distance between nearest neighbors is minimized at the center of the trap. Also show the laser wavevector ${\bf k}_{\rm L}$ within the $xy$-plane. {\bf (b)} Vibrational frequencies $\omega_{n\alpha}$ for the radial modes $\alpha=x,y$ of a linear ion chain.  As $\kappa_x\to\kappa_{\rm c}$, the frequency of the lowest-energy mode tends to zero $\omega_{1x}\to 0$. Also shown a scheme of a laser beam, which is detuned with respect to the electronic transition such that $\omega_2-\omega_0$ is close to the blue-sideband resonance $\omega_2-\omega_0\approx\omega_{ny}$. The large anisotropy $\omega_y\gg\omega_x$ makes the coupling of this laser to the vibrational modes along  $x$ negligible.  }
\label{phonon_modes_fig}
\end{figure}

Let us recall that the atomic plasma consists of $N$ ions of charge $e$ and mass $m$ coupled by means of the Coulomb interaction and confined by an effective quadratic  potential 
 \begin{equation}
 \label{potential}
 V=\sum_{i=1}^N\sum_{\alpha=x,y,z}\frac{1}{2} m\omega_{\alpha}^2r_{i\alpha}^2+\frac{e^2}{2}\sum_{i\neq j}\frac{1}{|{\bf r}_i-{\bf r}_j|}.
 \end{equation}
 At low temperatures, the ions self-assemble in a crystalline structure due to the balance of the trapping force and the Coulomb repulsion, namely $\boldsymbol{\nabla}_{{\bf r}_{i}}V={\bf 0}$. The solution of this system of equations yields  the equilibrium positions ${\bf r}_{i}^0=l_z{\tilde{{\bf r}}_i^0}$, where $l_z=(e^2/m\omega_z^2)^{1/3}$ is a unit of length, and determines the particular geometry of  the ion crystal. In this Appendix, we consider the regime of a linear ion crystal [Fig.~\ref{phonon_modes_fig}{\bf (a)}], and expand the potential~\eqref{potential}   around the equilibrium configuration ${\bf r}_i=l_z(\tilde{z}_i^0{\bf e}_z+q_ {ix}{\bf e}_x+q_ {iy}{\bf e}_y+q_ {iz}{\bf e}_z)$, considering  oscillations of a small amplitude $q_{i\alpha}\ll \tilde{z}_i^0$. The Taylor-expanded  potential becomes $V\approx V_0+V_{\rm h}+V_{\rm a}$, where $V_0$ stands for the configurational energy of the linear chain, $V_{\rm h}$ represents the harmonic approximation containing quadratic terms, and $V_{\rm a}$ is the anharmonic contribution due to the remaining non-linearities. The harmonic part can be written as follows
   \begin{equation}
   \begin{split}
 V_{\rm h}=\sum_{\alpha}\sum_{i,j}\frac{1}{2}m\omega_{\alpha}^2l_z^2\mathcal{V}_{ij}^{\alpha}q_{i\alpha}q_{j\alpha}
 \end{split}
 \end{equation} 
where  we have introduced the vibrational couplings 
\begin{equation}
\mathcal{V}_{ij}^{\alpha}=\left(1-c_{\alpha}\kappa_{\alpha}\sum_{l\neq i}\frac{1}{|\tilde{z}_i^0-\tilde{z}_l^0|^{3}}\right)\delta_{ij}+\frac{c_{\alpha}\kappa_{\alpha}}{|\tilde{z}_i^0-\tilde{z}_j^0|^{3}}(1-\delta_{ij}),
\end{equation}
which depend on  the anisotropy parameters $\kappa_{\alpha}=(\omega_{z}/\omega_{\alpha})^2$, $c_{\alpha}=1-3\delta_{\alpha z}$, and $\delta_{mn}$ stands for the Kronecker delta. The corresponding  Hamiltonian $H=\sum_{i\alpha} (l_z^{2}/2m)p_{i\alpha}^2+V_{\rm h}$  can be diagonalized by a canonical transformation (see e.g.~\cite{feynman_book}), 
\begin{equation}
Q_{n\alpha}=\sum_i\mathcal{M}_{in}^{\alpha}q_{i\alpha}, \hspace{1ex} P_{n\alpha}=\sum_i\mathcal{M}_{in}^{\alpha}p_{i\alpha},
\end{equation}
where the orthogonal matrices $\mathcal{M}^{\alpha}_{in}$ fulfill 
\begin{equation}
\sum_{ij}\mathcal{M}^{\alpha}_{in}\mathcal{V}^{\alpha}_{ij}\mathcal{M}^{\alpha}_{jm}=\mathcal{V}^{\alpha}_n\delta_{nm},\hspace{2ex} \sum_{n}\mathcal{M}^{\alpha}_{in}\mathcal{M}^{\alpha}_{jn}=\delta_{ij},
\end{equation}
such that $\mathcal{V}_n^{\alpha}$ yield the normal-mode frequencies $\omega_{n\alpha}=\omega_{\alpha}\sqrt{\mathcal{V}_n^{\alpha}}$  displayed  in Fig.~\ref{phonon_modes_fig}{\bf (b)}. In this figure, we represent the radial phonon modes  for a chain of $N=30$ ions, such that $\omega_y=30\omega_z$ and $\omega_x$ is varied. According to these normal modes, the harmonic Hamiltonian can be expressed as follows $H=\sum_{n\alpha}((l_z^2/2m)P_{n\alpha}^2+\half r_{n}^{\alpha}Q_{n\alpha}^2),$ where we have introduced the relevant coupling strength
\begin{equation}
\label{r}
r_{n}^{\alpha}=m\omega_{\alpha}^2l_z^2\left(1-\frac{c_{\alpha}\kappa_{\alpha}}{2}\sum_i\sum_{j\neq i}\frac{(\mathcal{M}_{in}^{\alpha}-\mathcal{M}_{jn}^{\alpha})^2}{|\tilde{z}_i^0-\tilde{z}_j^0|^3}\right).
\end{equation}
Note that for the so-called axial modes $c_z=-2$, this parameter   is constrained to be positive $r_{n}^{z}>0$. The same occurs for the radial modes along the strongly-confined $y$-axis. Even if $c_y=1$  in this case, the vanishingly small anisotropy $\kappa_y\ll 1$ leads to $r_{n}^{y}>0$. In both cases, the additional non-linearities of $V_{\rm a}$ can be neglected, and the  vibrational Hamiltonians correspond to a collection of uncoupled harmonic oscillators 
\begin{equation}
\label{phonon_modes}
H_y=\sum_n\omega_{ny}\left(a_{n}^{\dagger}a_{n}^{\phantom{\dagger}}+\frac{1}{2}\right),
\hspace{1ex}H_z=\sum_n\omega_{nz}\left(b_{n}^{\dagger}b_{n}^{\phantom{\dagger}}+\frac{1}{2}\right),
\end{equation}
where $a_n,b_n$ are the phonon annihilation operators.

\begin{figure}

\centering
\includegraphics[width=1\columnwidth]{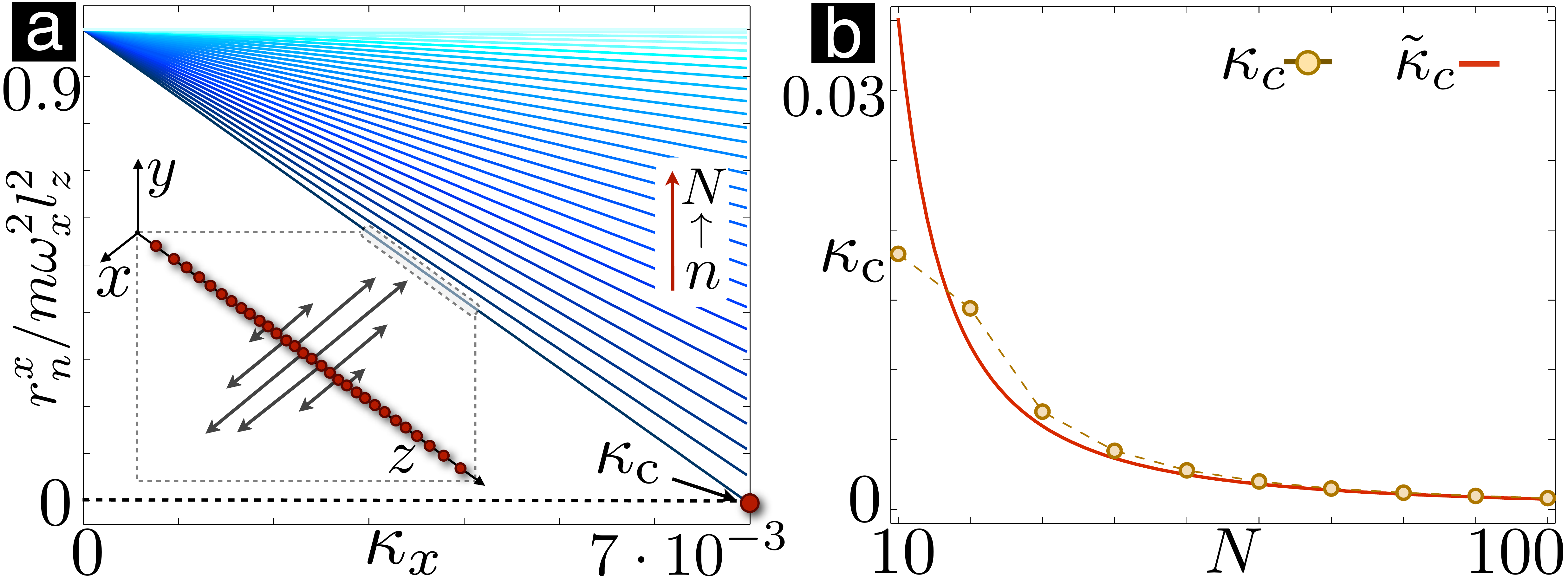}
\caption{ {\bf Mode softening for a linear ion chain:} {\bf (a)}   Parameters $r_n^x$ as a function of the anisotropy $\kappa_x$ for a linear Coulomb crystal with $N=30$ ions. As $\kappa_x\to\kappa_{\rm c}=7\cdot 10^{-3}$,  $r_{n_{\rm s}}^x\to 0$ for the lowest frequency mode, which corresponds to the zigzag distortion (see the mode $\mathcal{M}_{in_{\rm s}}$ in inset for $\kappa_{x}=\kappa_{\rm c}/2$). {\bf (b)} Scaling of the critical point $\kappa_{\rm c}$ as a function of the number of trapped ions, which follows from the solution of  Eq.~\eqref{r}  (yellow dots). This parameter is compared to the estimate   $\tilde{\kappa}_{\rm c}=64(\log(3Nx_1/2^{3/2})-1)/9N^2x_1$ (red solid line)~\cite{structural_transition}, where $x_1=(7\zeta(3)/2)^{1/2}$, and $\zeta(s)$ is the Riemann zeta function. }
\label{phonon_cond}
\end{figure}

A different situation occurs for the radial modes along the  $x$-axis, where $c_{x}=1$, and $\kappa_x\leq 1$ is not that small. In this case, there might exist a certain regime where $r_{n_{\rm s}}^x\to 0$ for a particular vibrational mode $n_{\rm s}\in\{1\cdots N\}$. This gapless vibrational mode,  the so-called soft mode, is the paradigm in the Landau theory of SPTs~\cite{spt_cowley}. As occurs in the homogeneous case~\cite{landau_theory}, the soft mode for the inhomogeneous linear-to-zizag transition corresponds to the zigzag mode $n_{\rm s}=1$,  which becomes gapless a the critical point $r_{n_{\rm s}}^x=0$ for $\kappa_x=\kappa_c$ [Fig.~\ref{phonon_cond}{\bf (a)}]. In this situation, one is obliged to consider the anharmonic terms 
\begin{equation}
V_{\rm a}=\sum_{i\neq j}\frac{3}{16}m\omega_z^2l_z^2\frac{1}{|\tilde{z}_i^0-\tilde{z}_j^0|^5}\left(\sum_n\mathcal{M}^x_{in}Q_{nx}-\sum_{m}\mathcal{M}_{jm}^xQ_{mx}\right)^4,
\end{equation} 
which couple the soft phonon  mode to the rest of the vibrational modes $n\neq n_{\rm s}$,  leading thus to a complicated many-body problem. However, the higher-energy modes can be  integrated out, yielding a set of renormalized parameters for the soft mode that incorporate the effects of  finite temperatures~\cite{rg_zigzag}. Since we are interested in the $T=0$ case, we neglect these corrections, and simply consider the effect of the quartic term on the soft zizag mode
\begin{equation}
\label{u}
V_{\rm a}=\frac{1}{4}u_{n_{\rm s}}^xQ_{n_{\rm s}x}^4,\hspace{1.5ex}u_{n_{\rm s}}^x=\frac{3}{4}m\omega_z^2l_z^2\sum_{i}\sum_{j\neq i}\frac{(\mathcal{M}^x_{in_{\rm s}}-\mathcal{M}_{jn_{\rm s}}^x)^4}{|\tilde{z}_i^0-\tilde{z}_j^0|^5}.
\end{equation} 
From this expression, it is evident that $u_{n_{\rm s}}^x>0$, which guarantees the stability of  the theory for the soft mode in the regime $r_{n_{\rm s}}^x<0$.  The effective description of the SPT corresponds to the so-called $\phi^4$ scalar field theory
\begin{equation}
H_{\rm s}=\frac{l_z^2}{2m}P_{n_{\rm s}}^2+\half r_{n_{\rm s}}^{x}Q_{n_{\rm s}x}^2+\frac{1}{4} u_{n_{\rm s}}^{x}Q_{n_{\rm s}x}^4,
\end{equation}
such that the broken symmetry phase $r_{n_{\rm s}}^x<0,u_{n_{\rm s}}^x>0$ yields the zigzag configurations of the ion crystal.
Note that, as customary in bosonic interacting models of this type, this broken-symmetry phase can be understood as a consequence of the macroscopic occupation of the soft mode $\langle Q_{n_{\rm s},x}\rangle\neq 0$ (i.e. phonon condensation). Two comments are now in order. Firstly, since the model corresponds to a real scalar field, this phonon condensation is not associated to the phenomenon of superfluidity familiar to trapped gases of weakly-interacting  bosonic atoms. Secondly, the usual long-range order associated to the condensed state only takes place at $T=0$, so this state of matter is rather referred to as a quasi-condensate.

Finally, let us address the accuracy of this description by comparing its predictions  to the estimates for inhomogeneous Coulomb crystals in~\cite{structural_transition}, which are based on a different approach based on the theory of  prolate spheroidal  plasmas. The critical point implicitly given by Eq.~\eqref{r} is
\begin{equation}
\label{cp}
\kappa_c^{-1}=\frac{1}{2}\sum_{i}\sum_{j\neq i}\frac{(\mathcal{M}_{in}^{x}-\mathcal{M}_{jn}^{x})^2}{|\tilde{z}_i^0-\tilde{z}_j^0|^3},
\end{equation}
which has been represented in Fig.~\ref{phonon_cond}{\bf (b)}, and compared to the analytical estimate in~\cite{structural_transition}. This critical point marks the onset of the linear-to-zigzag transition, which takes place at the center of the trap and progressively follows to the edges. The agreement of both predictions justifies the validity of the results presented in this Appendix.

 \subsection{Inhomogeneous structural phase transition and coupled double-well oscillators}
 \label{spt}
 
 In the previous Appendix, we discussed the onset of the linear-to-zigzag SPT for inhomogeneous ion crystals. The theory is based on the soft zigzag mode described by an effective $\phi^4$ model, which pinpoints the origin of the instability at the center of the trap, but does not account for its propagation towards the edges.  In order to describe this effect, we should resort to the local ion vibrations rather than to the collective phonon modes, but exploiting our  knowledge about the zigzag distortion being responsible of  the SPT. This can be achieved by the ansatz  $q_ {ix}=(-1)^i\delta q_ {ix}$ introduced in~\cite{ising_theory} for the homogeneous case, which we use here for the inhomogeneous crystal. Note that $\delta q_{ix}$ is a slowly varying transverse displacement, which  allows a gradient expansion $\delta q_{jx}\approx \delta q_{ix}+(\tilde{z}_j^0-\tilde{z}_i^0)\partial_{i+1}\delta q_{ix}$, where $\partial_{j}\delta q_{ix}=(\delta q_{ix}-\delta q_{jx})\ll\delta q_{ix}$. Note that this ansatz is inspired by the actual shape of the zigzag mode (see inset of Fig.~\ref{phonon_cond}{\bf (a)}), which corresponds to an alternating distortion modulated by a slowly varying envelope. Let us also remark that the above gradient expansion is only valid away from the edges, where the inhomogeneities become much larger, and higher powers of the gradient might be required. However, it suffices to describe  how the instability proceeds towards the edges. The Taylor expansion of the Coulomb interaction gives rise to 
  \begin{equation}
 \label{oscillators_x}
 \begin{split}
 V_{x}&=\sum_i\left(\frac{r^{x}_i}{2}\delta q_{i x}^2+\frac{1}{4}u^{x}_i\delta q_{i x}^4\right)+\sum_{i\neq j}\frac{K^{x}_{ij}}{2}(\partial_j\delta q_{i x})^2+\Delta V_x,
\end{split}
 \end{equation}
where $\Delta V_x$ accounts for the effect of the remaining vibrational modes.  Here, we have introduced the characteristic couplings of Eq. (4) in the main text, which depend upon the following function
\begin{equation}
\zeta_i(n)=\sum_{l\neq i}\frac{[(-1)^i-(-1)^l]^{n-1}}{|\tilde{z}_i^0-\tilde{z}_l^0|^{n}},
\end{equation}
where $n\in\mathbb{Z}$. Neglecting the effects of the remaining vibrational modes, the Hamiltonian $H_x=\sum_i (l_z^2/2m)p^2_{ix}+V_x$ corresponds to a set of anharmonic oscillators with site-dependent couplings
\begin{equation}
\label{r_app}
r^x_{i}=m\omega_x^2l_z^2\left(1-\half\kappa_x\zeta_i(3)\right),\hspace{1ex} u_i^x=m\omega_x^2l_z^2\left(\frac{3}{4}\kappa_x\zeta_i(5)\right),
\end{equation}
which clearly resemble the parameters of the zigzag $\phi^4$ model in Eqs.~\eqref{r} and~\eqref{u}. In addition, one finds a nearest-neighbor coupling strength
\begin{equation}
\label{k_app}
K_{ij}^x=m\omega_x^2l_z^2\left(\sum_{l\neq i}\frac{(-1)^{l+i+1}\kappa_x}{2|\tilde{z}_i^0-\tilde{z}_l^0|}\right)\delta_{j,i+1},
\end{equation}
leading to a collection of coupled double-well oscillators.  

\begin{figure}

\centering
\includegraphics[width=1\columnwidth]{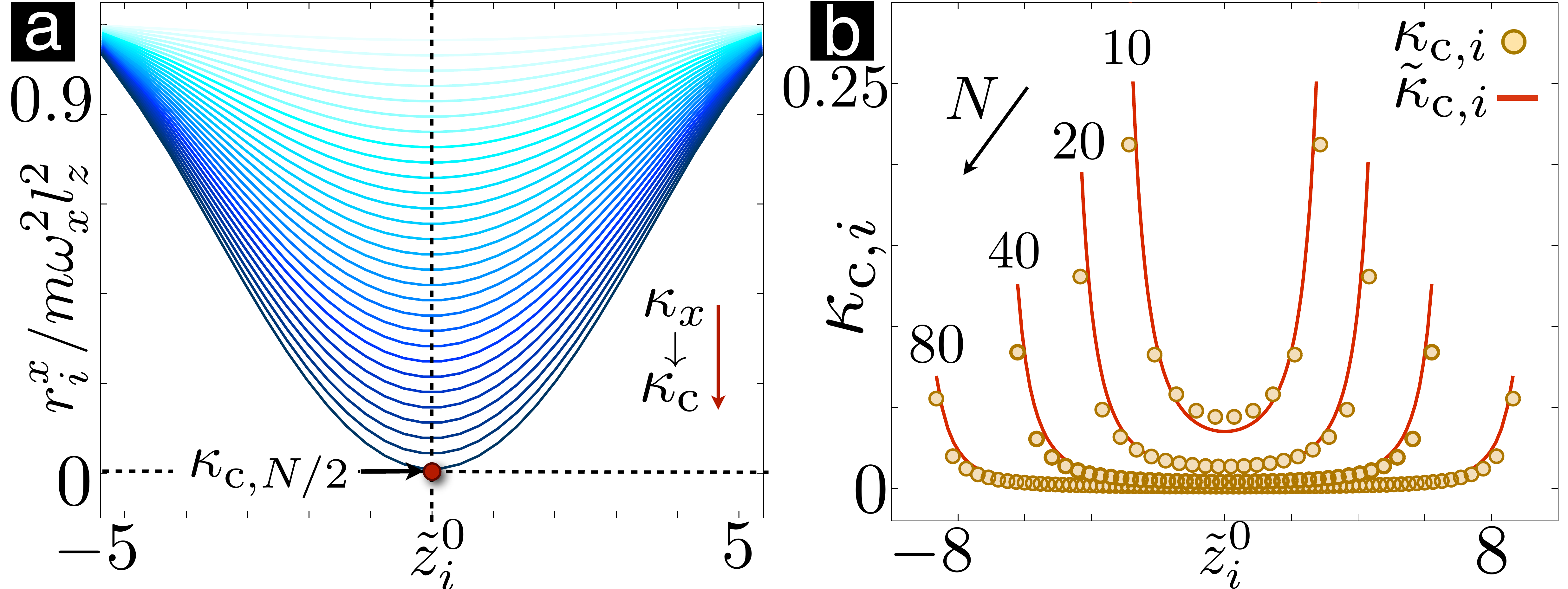}
\caption{ {\bf Local double-well structural phase transition:} {\bf (a)}    Parameter $r_i^x$ as a function of the ion position for different anisotropies  $\kappa_x$ in a linear Coulomb crystal with $N=30$ ions. As $\kappa_x\to\kappa_{{\rm c},N/2}$,  $r^x_{N/2}\to 0$, signaling the onset of the SPT at the center of the trap. {\bf (b)}  Comparison of the inhomogeneous critical point  $\kappa_{{\rm c},i}$  given by  Eq.~\eqref{cp_inhomogeneous}  (yellow dots), with   the analytical estimate~\cite{structural_transition}  that predicts $\tilde{\kappa}_{{\rm c},i}=\tilde{\kappa}_{\rm c}/(1-\Delta\tilde{\kappa}(\tilde{z}_i^0)^2)^3$, with $\Delta\tilde{\kappa}=16/(9N^2\tilde{\kappa}_{\rm c}^{2/3}x_1^{4/3})$ (red  line). We show the clear agreement between both predictions.}  
\label{phonon_cond_loc}
\end{figure}

By numerically evaluating Eq.~\eqref{k_app}, one finds that the oscillator coupling fulfills $u_i^x>0,K_{ij}^x>0$, and thus the lower energy configuration tries to minimize the gradient. Assuming that $\langle\delta q_{jx}\rangle=\langle\delta q_{ix}\rangle$ in the bulk of the ion chain, the broken-symmetry phase $\langle \delta q_{ix}\rangle\neq 0$  occurs when $r_{N/2}^x<0$ leading to the double-well potential. However, the site-dependent couplings induce an inhomogeneous nucleation that originates at the center of the trap [Fig.~\ref{phonon_cond_loc}{\bf (a)}]. The  inhomogeneous critical point is given by $\kappa_{{\rm c},i}=2/\zeta_i(3)$, namely
\begin{equation}
\label{cp_inhomogeneous}
\kappa_{{\rm c},i}^{-1}=\frac{1}{2}\sum_{j\neq i}\frac{[(-1)^i-(-1)^j]^2}{|\tilde{z}_i^0-\tilde{z}_j^0|^3},
\end{equation}
which increases as one proceeds to the edges of the ion crystal.  Let us remark that such an inhomogeneous critical point (yellow dots) has been compared to the analytical estimate based on spheroidal plasmas (red solid line)~\cite{structural_transition} in Fig.~\ref{phonon_cond_loc}{\bf (b)}. The clear agreement  justifies the validity of the above ansatz.

\begin{figure}

\centering
\includegraphics[width=1\columnwidth]{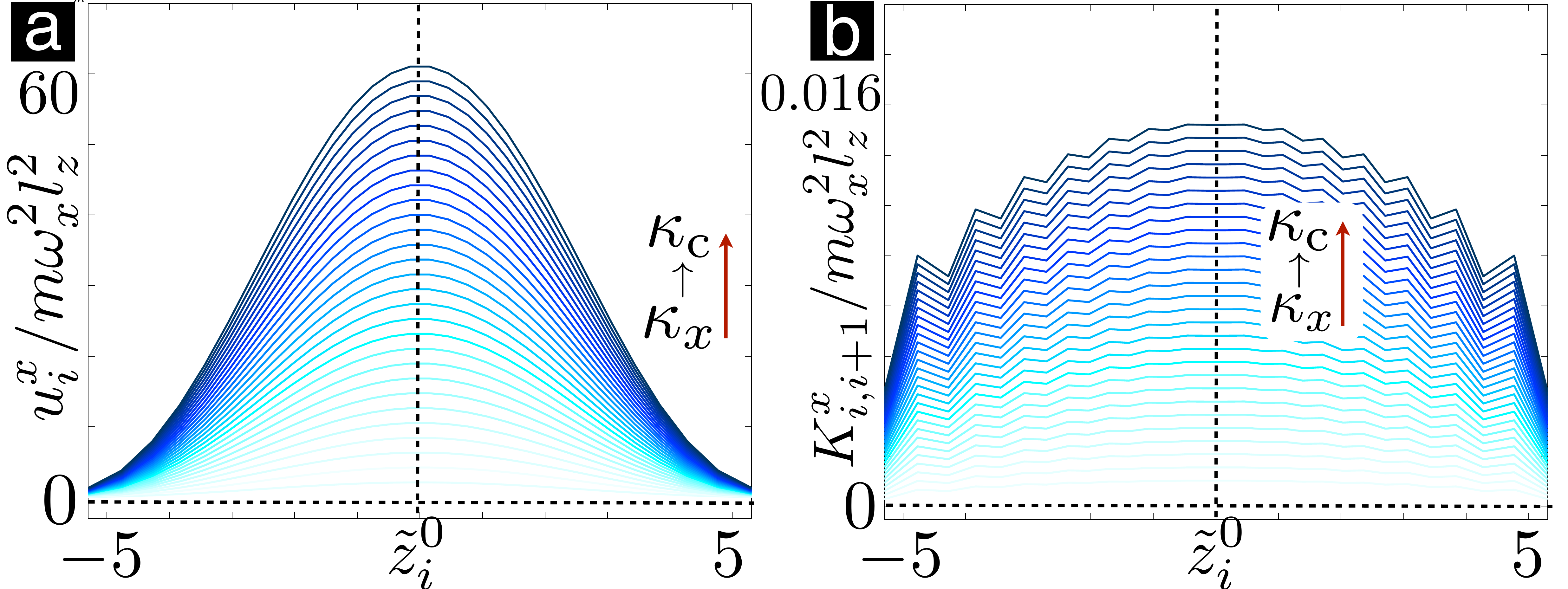}
\caption{ {\bf Inhomogeneous parameters:} {\bf (a)}    Parameter $u_i^x$ as a function of the ion position for different anisotropies  $\kappa_x$ in a linear Coulomb crystal with $N=30$ ions.  {\bf (b)}  Parameter $K_{i,i+1}^x$.}  
\label{loc_parameters}
\end{figure}

We have evaluated numerically the remaining parameters of Eqs.~\eqref{r_app} and~\eqref{k_app}. In  Fig.~\ref{loc_parameters}{\bf (a)}, we represent the parameter associated to the quartic contribution, which becomes more important in the vicinity of the critical point. Note that, although $\delta q_{ix}^4\ll \delta q_{ix}^2$, the quartic contribution becomes relevant when $\kappa_x\to\kappa_{\rm c}$ where $u_{ix}\gg r_{ix}$. Let us also note that the oscillator couplings $K_{ij}^x$ also become more relevant close to criticality (see Fig.~\ref{loc_parameters}{\bf (b)}). However, they are several orders of magnitude smaller than the remaining parameters. This justifies our treatment of the spin-Peierls phenomena, whereby we have neglected their effect in  a first approximation.

\subsection{Generalized XY model with Dipolar Interactions}

In this Appendix, we present a detailed derivation of the dimerized quantum Ising model (DQIM) responsible of the spin-Peierls transition. Let us start from the laser-ion interaction in Eq.(2) of the main text, which we rewrite here for convenience
   \begin{equation}
   \label{laser_ion_supp}
   H_{\rm L}=\frac{1}{2}\sum_{l,i}(\Omega_l\sigma_i^++\Omega_l^*\sigma_i^-)(\ee^{\ii{\bf k}_l\cdot{\bf r}_i-\ii\omega_lt+\ii\phi_l}+\ee^{-\ii{\bf k}_l\cdot{\bf r}_i+\ii\omega_lt-\ii\phi_l}).
   \end{equation}
We now express the position of the ions in terms of the collective phonon modes~\eqref{phonon_modes}, namely ${\bf r}_i=l_z\tilde{z}_i^0{\bf e}_z+\delta {{\bf r}}_i$, where
\begin{equation}
\delta {{\bf r}}_i=l_zq_ {ix}{\bf e}_x+\sum_n\frac{{\bf e}_y}{\sqrt{2m\omega_{ny}}}(a_n^{\phantom{\dagger}}+a_n^{\dagger})+\sum_n\frac{{\bf e}_z}{\sqrt{2m\omega_{nz}}}(b_n^{\phantom{\dagger}}+b_n^{\dagger}).
\end{equation}
As argued in previous sections, in order to treat the inhomogeneous SPT,  it is more appropriate to consider the vibrations along the $x$-axis in the local basis $q_ {ix}$. Note that  the laser wavevectors are aligned within the $xy$-plane [Fig.~\ref{phonon_modes_fig}{\bf (a)}], so that the spins are not coupled to the axial phonons $b_n,b_n^{\dagger}$, which shall be  neglected.  Besides, the large difference between the radial trapping frequencies, $\omega_x\ll\omega_y$, implies that the laser beams in resonance with the vibrational sidebands along the $y$-axis will be out of resonance with respect to the sidebands along the $x$-axis [Fig.~\ref{phonon_modes_fig}{\bf (b)}].  Note that the resonance with higher sidebands can also be neglected by controlling the trapping frequency $\omega_{x}$ and the orientation of the laser wavevector ${\bf k}_{\rm L}$, so that  the coupling of the spins to the vibrational modes along the $x$-axis can be ignored. There is however one exception, the soft phonons can condense in the zigzag mode  as a consequence of the SPT, affecting thus the laser-ion Hamiltonian. Hence, we must  consider  $H_0=\frac{1}{2}\sum_i\omega_0\sigma_i^z+\sum_n\omega_na_n^{\dagger}a_n^{\phantom{\dagger}}+H_{x}$, where  $\omega_n=\omega_{ny}$, and
 \begin{equation}
 H_{x}=\!\sum_i\!\left(\!\frac{ml_z^2}{2}\left(\partial_t \delta q_{ix}\right)^2\!+\frac{r^{x}_i}{2}\delta q_{ix}^2\!+\frac{u^{x}_i}{4}\delta q_{ix}^4\!\right)\!+\!\sum_{i\neq j}\!\frac{K^{x}_{ij}}{2}(\partial_j\delta q_{ix})^2\!,
 \end{equation}
according to Eqs.~\eqref{r_app} and~\eqref{k_app}. We now transform the laser-ion Hamiltonian to an interacting picture with respect to $H'_0=\sum_i\frac{\omega_0}{2}\sigma_i^z+\sum_n\omega_na_n^{\dagger}a_n^{\phantom{\dagger}}$. After a Taylor expansion of the Hamiltonian~\eqref{laser_ion_supp} for small Lamb-Dicke parameters $\eta_{ln}={ k}_{ly}/\sqrt{2m\omega_n}\ll1$, we neglect the fast-oscillating terms from the resulting spin-phonon Hamiltonian by using a rotating-wave approximation (RWA). This RWA  holds when the laser beams are tuned to the red  $\omega_1\approx\omega_0-\omega_y$, and blue $\omega_2\approx\omega_0+\omega_y$ sidebands, and fulfill the following constraints $\Omega_l\ll\omega_0,$
\begin{equation}
\eta_{1n}\Omega_1\ll|\omega_1-(\omega_0+\omega_n)|, \hspace{0.5ex}\eta_{2n}\Omega_2\ll|\omega_2-(\omega_0-\omega_n)|.
\end{equation}
 In this regime, the laser-ion Hamiltonian becomes a combination of the red- and blue-sideband spin-phonon couplings
   \begin{equation}
   H_{\rm L}=\sum_{in}\left(\mathcal{F}_{in}^{\rm r}\sigma_i^+a_n^{\phantom{\dagger}}\ee^{\ii \theta_{\rm r}q_{ix}-\ii\delta_{n{\rm r}}t}+\mathcal{F}_{in}^{\rm b}\sigma_i^+a_n^{\dagger}\ee^{\ii \theta_{\rm b}q_{ix}-\ii\delta_{n{\rm b}}t}+\text{H.c.}\right),
   \end{equation}
   where we have introduced the sideband coupling strengths $\mathcal{F}_{in}^{\rm r}=\frac{\ii}{2}\Omega_1\eta_{1n}\mathcal{M}_{in}\ee^{\ii\phi_1},\mathcal{F}_{in}^{\rm b}=\frac{\ii}{2}\Omega_2\eta_{2n}\mathcal{M}_{in}\ee^{\ii\phi_2}$, and the detunings $\delta_{n{\rm r}}=\omega_1-(\omega_0-\omega_n),\delta_{n{\rm b}}=\omega_2-(\omega_0+\omega_n)$. Let us remark that in this spin-phonon Hamiltonian, the zigzag displacement that pinpoints the structural phase transition of the ion crystal is also considered in $q_{ix}=(-1)^i\delta q_{ix}$, such that the commensurability parameters are $\theta_{\rm r}=k_{1x}l_z,\theta_{\rm b}=k_{2x}l_z$. By making the sideband detunings opposite to each other $\delta_{n{\rm r}}=-\delta_{n{\rm b}}=:\delta_n$, it is possible to move to a picture where the phonons absorb the time-dependence, and the laser-ion Hamiltonian becomes 
  \begin{equation}
  \label{spin_phonon}
   H'_{\rm L}=\sum_{n}\delta_na_n^{\dagger}a_n^{\phantom{\dagger}}+\sum_{in}\left(\mathcal{S}_{in}^{\dagger}a_n^{\phantom{\dagger}}+\text{H.c.}\right),
  \end{equation}
where we have introduced the following  spin operator 
\begin{equation}
\label{spin_op}
\mathcal{S}_{in}=(\mathcal{F}_{in}^{\rm r})^*\sigma_i^-\ee^{-\ii\theta_{\rm r}q _{ix}}+\mathcal{F}_{in}^{\rm b}\sigma_i^+\ee^{+\ii\theta_{\rm b}q_{ix}}.
\end{equation}
 The current formulation of this generalized spin-phonon Hamiltonian~\eqref{spin_phonon} is amenable of performing perturbation theory in the regime $|\mathcal{F}_{in}^{\rm r/b}|\ll\delta_n$. Note that in this limit,  the phonons along the $y$-axis only get virtually excited and it is possible to integrate them out to obtain an effective Hamiltonian that only involves the spin degrees of freedom. Such type of perturbative expansion can be performed systematically by a polaron-type transformation 
\begin{equation}
U=\ee^{S},\hspace{1ex}S=\sum_{in}\left(\frac{1}{\delta_n}\mathcal{S}_{in}^{\phantom{\dagger}}a^{\dagger}_n-\frac{1}{\delta_n}\mathcal{S}^{\dagger}_{in}a^{\phantom{\dagger}}_n\right).
\end{equation}
At this point, we perform a mean-field approximation for the zigzag mode, so that $q_{ix}=\langle q_{ix}\rangle$ can be treated as a $c$-number. The transformed Hamiltonian can be obtained to any order of perturbation theory by using the identity $\ee^{S}H_{\rm L}'\ee^{-S}=H_{\rm L}'+[S,H_{\rm L}']+\frac{1}{2!}[S,[S,H_{\rm L}']]+\cdots$. To lowest order, we find $UH_{\rm L}'U^{\dagger}\approx \sum_n\delta_na_n^{\dagger}a_n^{\phantom{\dagger}}+H_{\rm eff}$, where 
\begin{equation}
H_{\rm eff}=-\sum_{ij}\sum_n\frac{1}{\delta_n}\mathcal{S}_{in}^{{\dagger}}\mathcal{S}_{jn}^{\phantom{\dagger}}+\sum_{inm}\mathcal{{N}}_{inm}\sigma_i^z,
\end{equation}
and we have introduced the following phonon operator 
\begin{equation}
\mathcal{N}_{inm}=\frac{\mathcal{M}_{in}\mathcal{M}_{im}}{4\delta_m}\left(\Omega_2^2\eta_{n2}\eta_{m2}-\Omega_1^2\eta_{n1}\eta_{m1}\right)\left(a_m^{\dagger}a_n^{\phantom{\dagger}}+a_n^{\dagger}a_m^{\phantom{\dagger}}\right).
\end{equation}
From this last expression, it is clear that by setting the laser parameters according to the following constraint
\begin{equation}
\Omega_1^2k_{1y}^2=\Omega_{2}^2k_{2y}^2,
\end{equation}
the residual spin-phonon operator vanishes, and we obtain an effective model of interacting spins. In particular, this constraint is satisfied  for a pair of counter-propagating laser beams  ${\bf k}_1=-{\bf k}_2=:{\bf k}$ with identical Rabi frequencies $\Omega_1=\Omega_2=:\Omega_{\rm L}$. By substituting the generalized spin operators in~\eqref{spin_op}, we obtain the following effective spin Hamiltonian
\begin{equation}
H_{\rm eff}=\sum_{i\neq j}\left(J_{ij}^{xx}\sigma_i^x\sigma_j^x+J_{ij}^{yy}\sigma_i^y\sigma_j^y+J_{ij}^{xy}\sigma_i^x\sigma_j^y+J_{ij}^{yx}\sigma_i^y\sigma_j^x\right),
\end{equation}
where the coupling strengths are those written in Eq.~(8) of the main text. This Hamiltonian corresponds to a generalization of the famous XY model~\cite{xy_model}, where in addition to the well-studied anisotropy between the $J_{ij}^{xx}$ and $J_{ij}^{yy}$ couplings, there are additional crossed couplings $J_{ij}^{xy},J_{ij}^{yx}$ which have remained largely unexplored. These couplings are expressed in terms of   the relative phase between the lasers $\phi_-=\phi_{1}-\phi_2$, and the strength
\begin{equation}
J_{ij}=-\sum_n\frac{\Omega_{\rm L}^2}{16\delta_n}\eta_n\eta_m\mathcal{M}_{in}\mathcal{M}_{jm}.
\end{equation}
Note that in the limit of tight transverse confinement $\kappa_y=(\omega_z/\omega_y)^2\ll1$, the above couplings  can be expressed as
\begin{equation}
\label{dipolar_coupling}
J_{ij}=\frac{J_{\rm eff}}{2|{\bf \tilde{r}}_i^0-{\bf \tilde{r}}_j^0|^3},\hspace{1.5ex}J_{\rm eff}=\frac{\Omega_{\rm L}^2\eta_y^2\kappa_y}{16\delta_y^2}\left(1+\frac{\delta_y}{\omega_y}\right)\omega_y,
\end{equation}
where we have introduced the bare detuning $\delta_y=\omega_1-(\omega_0-\omega_y)=-\omega_2+(\omega_0+\omega_y)$. Therefore, the effective spin model is precisely the { generalized XY model with dipolar interactions} expressed in Eqs.~(8)-(9) of the main text.

 In order to obtain the final expression of the Hamiltonian, note that the laser wavevectors lie in the $xy$-plane and are slightly tilted with respect to the $y$-axis [Fig.~1{\bf (c)} main text]. Accordingly, we Taylor expand  the spin couplings for the small parameter $\xi_i=\theta\langle\delta q_{ix}\rangle\ll1$. By considering that the laser beams are phase locked $\phi_-=0$, we obtain the announced dimerized QIM
\begin{equation}
J_{ij}^{xy}=2J_{ij} (-1)^{i+1}=J_{ji}^{yx}.
\end{equation}

\subsection{Dimerized Quantum Ising Model}

\subsubsection{Analytical Solution}

In this part of the Appendix, we present the analytical solution for the energy spectrum of the DQIM in Eq.~(11) of the main text. By restricting  to nearest neighbors in a homogeneous  crystal, the Hamiltonian becomes
\begin{equation}
\label{dqim}
H_{\rm eff}=\sum_{i=1}^NJ\big(\sigma_i^x\sigma_{i+1}^x-(-1)^{i}\xi\sigma_i^{x}\sigma_{i+1}^y+(-1)^{i}\xi\sigma_i^y\sigma_{i+1}^x-g\sigma_i^z\big),
\end{equation}
where $J>0$ stands for the antiferromagnetic spin coupling, and $\xi, g>0$ represent the dimerization and transverse field expressed in units of the spin coupling. In contrast to previous studies~\cite{xy_alternating,spin_peierls_heisenberg},  the dimerization term in Eq.~\eqref{dqim} does not commute with the bare spin couplings and thus introduces additional quantum fluctuations. Motivated by the derivation of the Hamiltonian for a system of trapped ions presented in the previous section, we shall focus on the limit of small dimerization $\xi\ll1$, where the magnetic phases coincide with those of the usual QIM, namely, the paramagnetic (P) and the antiferromagnetic (AF) phases. As shown below, the dimerization is responsible of {\it (a) }lowering the ground-state energy, and {\it (b) } displacing the  critical point. This two-fold effect will be exploited to show that it is possible to observe a  { spin-Peierls quantum  phase transition}   with trapped ions.  

We diagonalize the Hamiltonian~\eqref{dqim}  via the so-called Jordan-Wigner (JW)  transformation~\cite{jw}. We note that the effect of alternating dimerized couplings in the XY model has been addressed in~\cite{xy_alternating}. To the best of our knowledge, the dimerized crossed couplings in Eq.~\eqref{dqim} have not been studied so far. The JW mapping expresses the spin operators  in terms of fermionic operators. The spin up (down) states are associated to the presence (absence) of a spinless fermion $\ket{\!\uparrow}_i\leftrightarrow c_i^{\dagger}\ket{\text{vac}}_i$, $\ket{\!\downarrow}_i\leftrightarrow\ket{\text{vac}}_i$, by a non-local transformation 
\begin{equation}
\sigma_i^{+}=c_i^{\dagger}\ee^{\ii\pi\sum_{j<i}c_j^{\dagger}c_{j}^{\phantom{\dagger}}},\hspace{1ex}\sigma_i^{-}=\ee^{-\ii\pi\sum_{j<i}c_j^{\dagger}c_{j}^{\phantom{\dagger}}}c_i^{\phantom{\dagger}},\hspace{1ex}\sigma_i^z=2c_i^{\dagger}c_i^{\phantom{\dagger}}-1,
\end{equation} 
where we have introduced $\sigma_i^{\pm}=(\sigma_i^x\pm\ii\sigma_i^y)/2$. The fermionized Hamiltonian can be expressed as follows
\begin{equation}
H_{\rm eff}=\sum_iJ\bigg(\big(1+\ii(-1)^{i}2\xi\big)c_{i+1}^{\dagger}c_i^{\phantom{\dagger}}-2gc_i^{\dagger}c_i^{\phantom{\dagger}}-c_{i+1}^{\dagger}c_{i}^{\dagger}\bigg)+\text{H.c.},
\end{equation}
which corresponds to a tight-binding model of spinless fermions with dimerized complex tunnelings $t=J(1\pm\ii2\xi)$, on-site energies $\epsilon=-2Jg$, and a pairing term between nearest neighbors that can be interpreted as the injection of Cooper pairs from a neighboring superconductor with gap  $\Delta=-J$. This Hamiltonian is diagonalized in momentum space $c_j=\sum_q\ee^{\ii qj}c_q$, where $q\in[-\pi,\pi)$ stands for the crystal momentum, and we have set the lattice spacing to unity. By introducing the Nambu spinor $\Psi(q)=(c_q,c_{-q}^{\dagger},c_{q-\pi},c^{\dagger}_{-q+\pi})^{t}$ in a reduced Brillouin zone, the Hamiltonian becomes 
\begin{equation}
H=\sum_{ q\in{\rm RBZ}}\Psi(q)^{\dagger}\mathbb{H}_q\Psi(q),\hspace{2ex} {\rm RBZ=[0,\pi/2)},
\end{equation} 
where  $\mathbb{H}_q$ is expressed in terms of  Pauli matrices
\begin{equation}
\mathbb{H}_q=2J\big(\cos q\tau_z\otimes\tau_z+\sin q\tau_z\otimes\tau_y-2\xi\cos q\tau_y\otimes\mathbb{I}_2-g\mathbb{I}_2\otimes\tau_z\big).
\end{equation}
The diagonalization of this matrix yields four eigenvalues $U_q\mathbb{H}_qU_q^{\dagger}={\rm diag}\{-\epsilon_{-}(q),-\epsilon_{+}(q),+\epsilon_{-}(q),+\epsilon_{+}(q)\}$, where 
\begin{equation}
\label{spectrum}
\epsilon_{\pm}(q)= 2J\sqrt{(g\pm\sqrt{1+4\xi^2}\cos q)^2+\sin^2q}.
\end{equation}
The associated unitary matrices define a generalized Bogoliubov transformation $(\gamma_{q,1},\gamma_{q,2},\gamma_{q,3},\gamma_{q,4})^t=U_q\Psi(q)$ that diagonalizes the fermionized  Hamiltonian. The groundstate of the system corresponds to all the negative-energy levels  filled $\ket{\rm g}=\prod_{q}\gamma_{q,1}^{\dagger}\gamma_{q,2}^{\dagger}\ket{\rm vac}$, where $\ket{\rm vac}$ stands for the vacuum. In the thermodynamic limit, the ground-state energy $E_{\rm g}^{\rm JW}=-\sum_{q\in{\rm RBZ}}(\epsilon_{-}(q)+\epsilon_{+}(q))$ can be expressed as follows
\begin{equation}
\label{gs_energy_supp}
E^{\rm JW}_{\rm g}(\xi)=-\frac{JN}{\pi}\int_0^{\pi}{\rm d}q\sqrt{(g-\sqrt{1+4\xi^2}\cos q)^2+\sin^2q}.
\end{equation}
To check the consistency of this result, note that  this energy coincides with the ground-state energy of the standard QIM~\cite{tfi} in the limit of vanishing dimerization $\xi=0$, namely $E^{\rm QIM}_{\rm g}(0)=-\frac{2}{\pi}JN(1+g){\mathcal{E}}(\theta_g)$. Here, we have introduced the complete elliptic integral of the second kind  
\begin{equation}
\mathcal{E}(\theta_{g})=\int_0^{\frac{\pi}{2}}{\rm d}\alpha(1-\theta_{g}^2\sin^2\alpha)^{1/2},
\end{equation}
where  $\theta_g=\sqrt{4g/(1+g)^2}$, and have used its properties listed in~\cite{tables_integrals}. With these expressions, we can address the effects of the dimerization announced at the beginning of this Appendix.

\begin{figure}

\centering
\includegraphics[width=1\columnwidth]{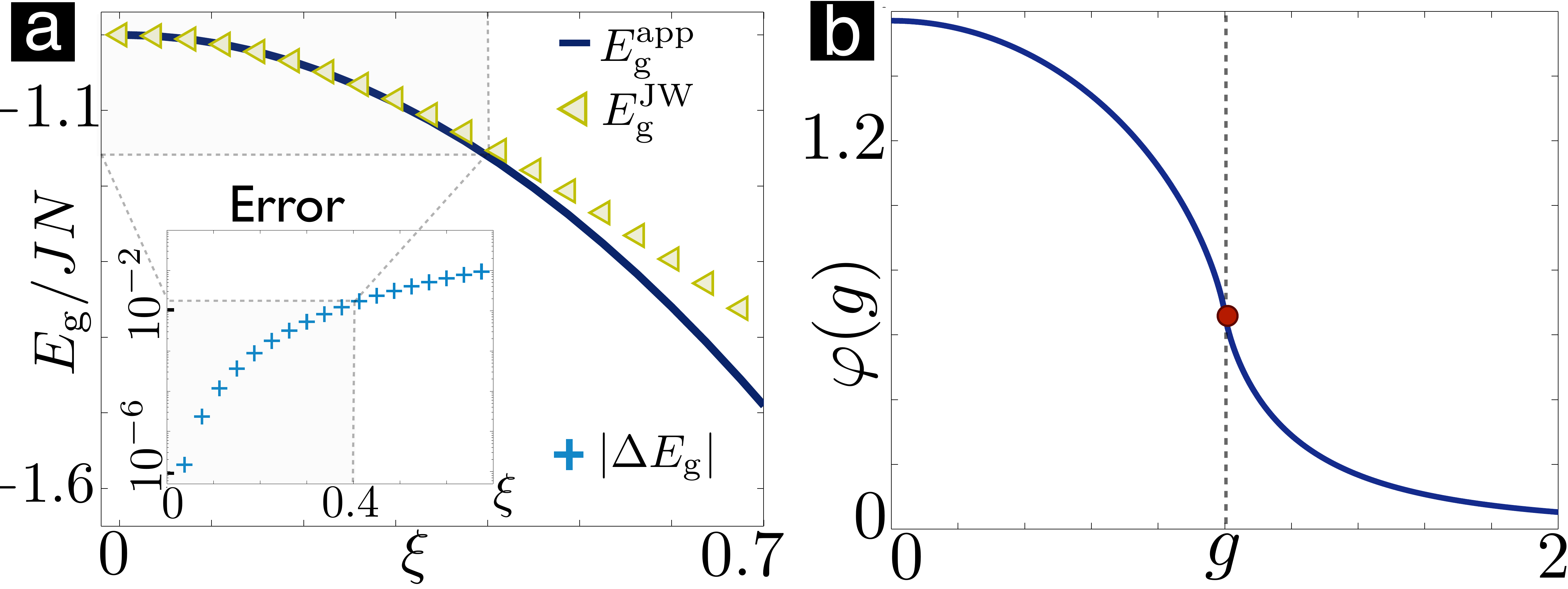}
\caption{ {\bf Scaling of the groundstate energy} {\bf (a)}    Groundstate energy  of the DQIM  as a function of the dimerization parameter $\xi$. The exact expression $E_{\rm g}^{\rm JW}(\xi)$~\eqref{gs_energy_supp} is represented by yellow triangles, and compared to the scaling $E_{\rm g}^{\rm app}(\xi)$~\eqref{gs_lowering} in blue solid lines In the inset, we display the error due to the approximation for $\xi\ll1$, namely $|\Delta E_{\rm g}|=|E_{\rm g}^{\rm JW}-E_{\rm g}^{\rm app}|$. {\bf (b)} The function $\varphi(g)$, which determines the quadratic scaling of the groundstate energy~\eqref{gs_lowering}, is a positive-definite and monotonically-decreasing function.}  
\label{dqim_qpt}
\end{figure}

 {\it (a)  Lowering of the ground-state energy.--} Considering the limit $\xi\ll1$ and the ground-state energy in Eq.~\eqref{gs_energy_supp}, we find
 \begin{equation}
 \label{gs_lowering}
 E^{\rm app}_{\rm g}(\xi)-E_{\rm g}^{\rm JW}(0)=\int_0^{\xi}{\rm d}\xi'\frac{\partial E^{\rm JW}_{\rm g}}{\partial \xi'}=-\frac{2JN}{\pi}\varphi(g)\xi^2,
 \end{equation}
 where we have introduced the following function 
 \begin{equation}
 \varphi(g)=\int_0^{\pi}{\rm d}q(\cos^2q-g\cos q)(1+g^2-2g\cos q)^{{\rm -}1/2}.
 \end{equation}
By using the complete elliptic integral of the first kind\begin{equation}
\mathcal{K}(\theta_{g})=\int_0^{\frac{\pi}{2}}{\rm d}\alpha(1-\theta_{\rm g}^2\sin^2\alpha)^{-1/2},
\end{equation}
this function can be expressed in a compact form
\begin{equation}
 \varphi(g)\!=\!\frac{1}{3g^2}\!\bigg(\mathcal{E}\!(\theta_g)(1+g)(2g^2-1)\!+\!\mathcal{K}\!(\theta_g)(1-g)(2g^2+1)\bigg).
\end{equation}
The above equation~\eqref{gs_lowering}  is precisely the expression~(12) used in the main text of this article. To test its validity,  we compare it to the exact expression $E^{\rm JW}_{\rm g}(\xi)$ in~\eqref{gs_energy_supp} for $g=0$, where $\varphi(0)=\pi/2$ (see Fig.~\ref{dqim_qpt}{\bf (a)}). As shown in the inset of this figure, the agreement between both expressions is extremely good even for considerable dimerization parameters $\xi\leq0.4 $. Since this regime contains the limit that can be attained with the ion trap QS, we can use directly the groundstate energy $E^{\rm app}_{\rm g}(\xi)$. From Fig.~\ref{dqim_qpt}{\bf (b)}, and from the properties of the elliptic integrals, $\varphi(g)$  is positive-definite function that gives rise to the aforementioned lowering of the groundstate energy  $E_{\rm g}^{\rm JW}(\xi)-E_{\rm g}^{\rm JW}(0)\approx-\frac{2JN}{\pi}|\varphi(g)|\xi^2<0$. This property becomes  crucial for the spin-Peierls transition, since it will compensate the energy increase due to the structural ion change.

 {\it (b)  Displacement of the quantum critical point.--} As mentioned at the beginning of this Appendix, the DQIM for $\xi\ll1$ encompasses two  magnetic phases that coincide with those of the usual QIM, namely, the paramagnetic and antiferromagnetic phases. The antiferromagnetic (AF) phase  has a degenerate groundstate that corresponds to the two antiparallel N\'eel configurations  $\ket{{\rm AF}}\in\{\ket{+-\cdots+-},\ket {-+\cdots-+}\}$ in the limit of $g\ll1$, where we have introduced the up/down spins in the $x$-basis $\ket{\pm}=(\ket{\!\uparrow}\pm\ket{\!\downarrow})/\sqrt{2}$. This degeneracy is related to the  invariance of the Hamiltonian under a global spin inversion $U_{\mathbb{Z}_2}=\prod_i\sigma_i^z$, which also applies to the DQIM.  Conversely, for $g\gg1$, the paramagnetic (P) phase has a unique groundstate where all the spins are parallel to the transverse field $\ket{{\rm P}}=\ket{\uparrow\uparrow\cdots\uparrow}$. As the transverse field is decreased, the paramagnetic groundstate evolves towards one of the possible antiferromagnetic states, breaking spontaneously  the symmetry at the quantum critical point $g=g_{\rm c}$. At this symmetry-breaking point, the energy gap between the groundstate and the lowest-lying excitations vanishes. 

 \begin{figure}

\centering
\includegraphics[width=1\columnwidth]{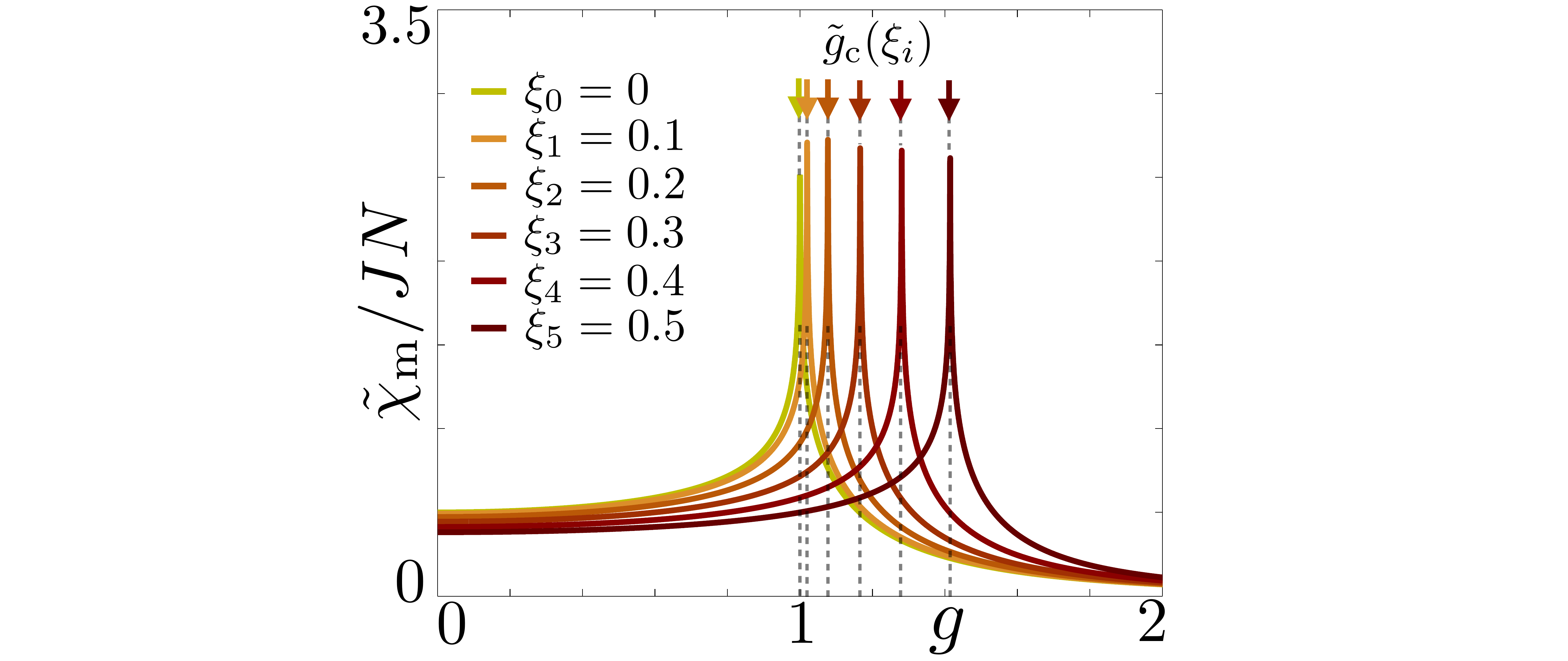}
\caption{ {\bf Displacement of the critical point:} Magnetic susceptibility $\chi_{\rm m}$ as a function of the transverse-field ratio $g$. The divergence of the susceptibility pinpoints the critical point $g_{\rm c}$ where the quantum phase transition from a paramagnet to an antiferromagnet takes place. The arrows indicate the flow of the critical point $g_{\rm c}\to\tilde{g}_{\rm c}(\xi)$ for different dimerization parameters $\xi_i\in\{0,0.1,0.2,0.3,0.4,0.5\}$}.  
\label{suscept}
\end{figure}

  For the QIM, the critical point is  located at $g_{\rm c}=1$, as imposed by the self-duality of the Hamiltonian under the order-disorder transformation~\cite{duality}, $\mu_i^z=\prod_{j<i}\sigma_i^z,\hspace{1ex} \mu_i^x=\sigma_{i-1}^x\sigma_i^x$, 
 \begin{equation}
  \label{duality}
  H(\{\sigma\},g)=-gH(\{\mu\},g^{-1})=-\sum Jg\big(\mu_i^x\mu_{i+1}^x-g^{-1}\mu_i^z\big),
 \end{equation} 
 which implies that the energy gap $\Delta(g_{\rm c})=0=\Delta(1/g_{\rm c})$ only vanishes at $g_{\rm c}=(g_{\rm c})^{-1}$, such that the critical point gets locked to $g_{\rm c}=1$. Note that this self-dual symmetry is explicitly broken by the dimerization~\eqref{dqim}, which lifts the constraint and allows the quantum critical point  to attain different values. In Fig.~\ref{suscept}, we represent the magnetic susceptibility $\chi_{\rm m}\propto\tilde{\chi}_{\rm m}=-\partial^2 E_g/\partial g^2$ as a function of $g$ for different dimerizations $\xi_i\in\{0,0.1,0.2,0.3,0.4,0.5\}$. The divergence of the susceptibility marks the critical point of the QPT. From this figure, it is evident that the critical point $g_{\rm c}=1$ flows towards higher values $\tilde{g}_{\rm c}(\xi)$ as the dimerization increases.

 In order to find analytically the location of the displaced critical point, we shall use the explicit expression for the energy bands~\eqref{spectrum}. The lowest-lying excitations occur at  $q=0$, such that their energy gap  scales as $\Delta \propto|g-\sqrt{1+4\xi^2}|$. The divergence of the correlation length typical of critical phenomena, which implies $\Delta(\tilde{g}_{\rm c})=0$, leads to
 \begin{equation}
 \label{mcp}
g_{\rm c}\to \tilde{g}_{\rm c}(\xi)=\sqrt{1+4\xi^2},
 \end{equation} 
 which is precisely the result in Eq.~(13) of the main text.
  Hence, the  critical point is displaced towards higher values of the transverse field. This effect will also  turn out to be essential for the quantum nature of the spin-Peierls transition.
  
 \subsubsection{Numerical Analysis}

 The increased complexity of the model brought by the dimerization demands a careful assessment  of the validity of the above results. In this part of the Appendix, we make a direct comparison between our analytical derivations and  the numerical results obtained by the density-matrix renormalization group (DMRG) method~\cite{dmrg_white}. DMRG algorithms are considered nowadays as the most efficient approach to one-dimensional QMBS, and therefore ideally suited to explore the strongly-correlated spin model~\eqref{dqim}. We focus on the groundstate properties of the model, and provide numerical evidence that clearly supports the validity of the expressions for the energy~\eqref{gs_energy_supp}, and its  scaling with the dimerization~\eqref{gs_lowering}.

\begin{figure}
\centering
\includegraphics[width=1\columnwidth]{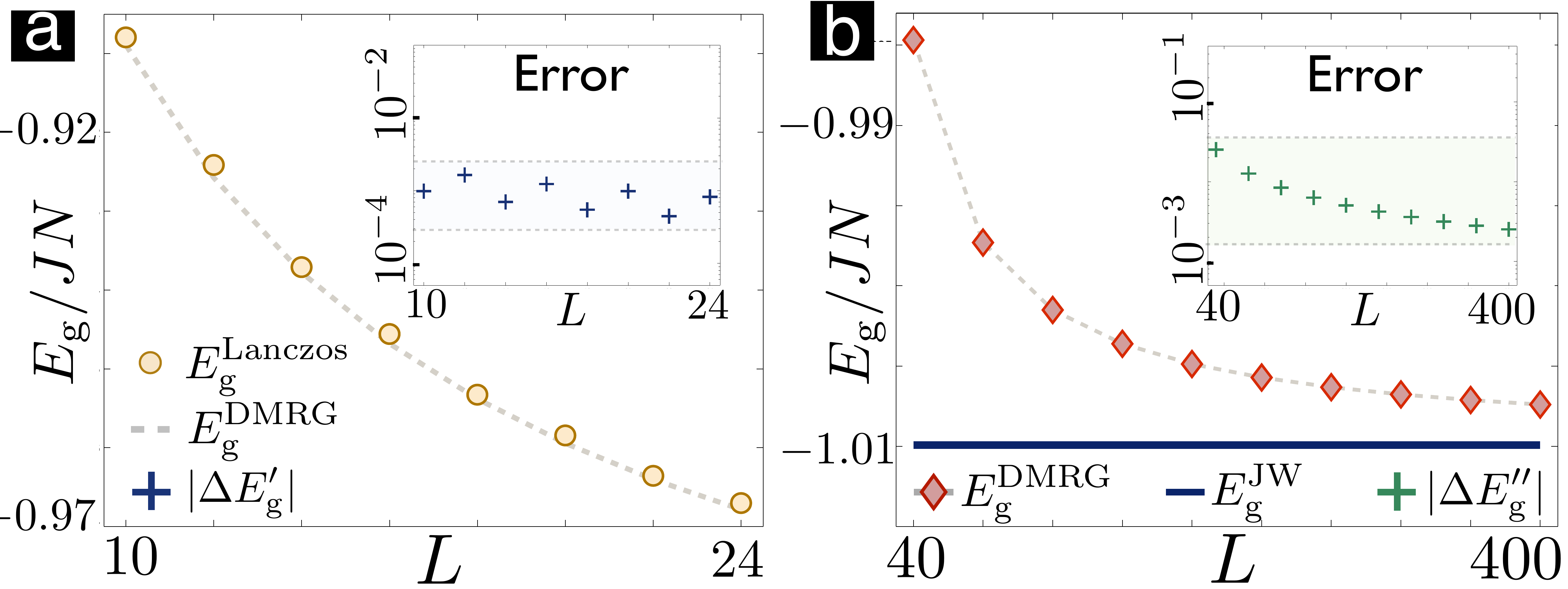}
\caption{ {\bf DMRG calculation of the groundstate energy:} {\bf (a)}    Comparison of the groundstate energy  of the DQIM  obtained by exact diagonalization methods based on a Lanczos algorithm  $E_{\rm g}^{\rm Lanczos}(\xi)$ (yellow dots), and the DMRG method $E_{\rm g}^{\rm DMRG}(\xi)$ (dashed grey line). We analyze $\xi=0.1,g=0$ for small chains with $L\leq 24$ sites. In the inset we represent the error $|\Delta E'_{\rm g}|=|E_{\rm g}^{\rm Lanczos}-E_{\rm g}^{\rm DMRG}|$ (blue crosses). {\bf (b)}  Comparison of the analytical estimate for the groundstate energy $E_{\rm g}^{\rm JW}(\xi)$ (blue solid line) in the  limit $L\to\infty$, and the  DMRG result $E_{\rm g}^{\rm DMRG}(\xi)$ (red diamonds) for longer chains of up to $L_{\rm max}=400$ sites, and $\xi=0.1,g=0$. In the inset we represent the error  $|\Delta E''_{\rm g}|=|E_{\rm g}^{\rm JW}-E_{\rm g}^{\rm DMRG}|$ (green crosses).}  
\label{dmrg}
\end{figure}

We use the finite-system DMRG algorithm for open boundary conditions. In order to test the accuracy of our algorithm, we have compared the obtained groundstate energies $E_{\rm g}^{\rm DMRG}$ to the results of a Lanczos algorithm $E_{\rm g}^{\rm Lanczos}$ for the exact diagonalization of small spin chains of length $L\in\{10,12,14,16,18,20,22,24\}$. Note that in order to achieve $L_{\rm max}=24$, we have optimized the Lanczos algorithm using the bipartite nature of the spin model~\eqref{dqim}. By dividing the chain in two blocks, it is possible to lower  the computational cost of matrix-vector operations, thus optimizing the performance of the Lanczos algorithm. For the DMRG algorithm, it suffices to keep $m=20$ states of the reduced block density matrices, and  considered $N_{\rm s}=1$  full sweeps of the renormalization procedure. In Fig.~\ref{dmrg}{\bf (a)}, we represent both energies for the DQIM~\eqref{dqim} after setting $\xi=0.1,g=0$. The good agreement serves as a testbed for our DMRG algorithm, which we now use to test  the analytical prediction~\eqref{gs_energy_supp} valid for the thermodynamical limit $L\to\infty$.

 Let us remark that the exact diagonalization cannot be carried far beyond $L_{\rm max}=24$. In clear contrast, the DMRG algorithm can be applied to longer chains, allowing us to consider $L\in\{40,80,120,160,200,240,280,320,360,400\}$. In Fig.~\ref{dmrg}{\bf (b)}, we represent the obtained energy $E_{\rm g}^{\rm DMRG}$, and compare it to the Jordan-Wigner estimate  $E_{\rm g}^{\rm JW}$~\eqref{gs_energy_supp} for  $\xi=0.1,g=0$. In this case we have set $m=50$ and $N_{\rm s}=2$. Note that  the analytical prediction gets more accurate as the thermodynamical limit is approached (see the inset of Fig.~\ref{dmrg}{\bf (b)}).
 
\begin{figure}

\centering
\includegraphics[width=1\columnwidth]{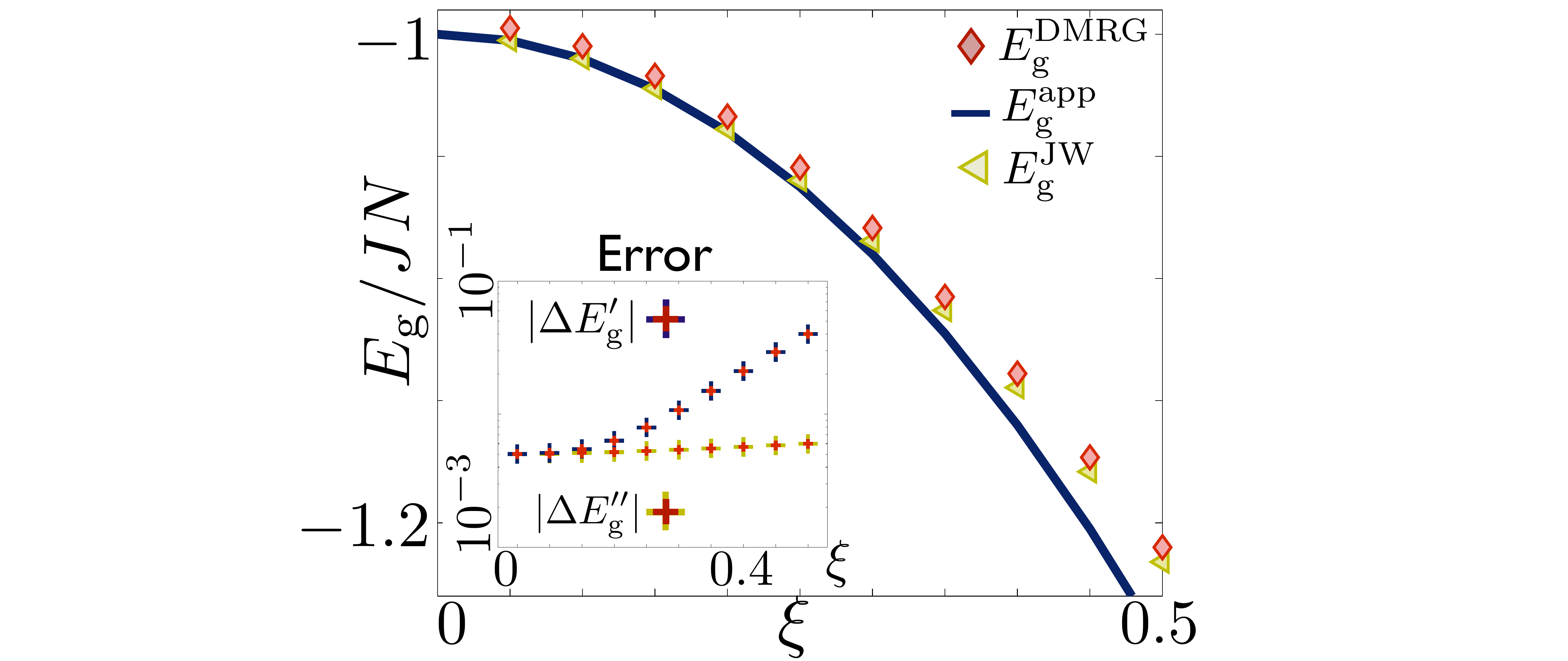}
\caption{ {\bf DMRG energy scaling law:} Comparison of the groundstate energy  obtained by the DMRG method $E_{\rm g}^{\rm DMRG}(\xi)$ (red diamonds) with the Jordan-Wigner solution $E_{\rm g}^{\rm JW}(\xi)$ (yellow triangles), and the approximate expression  $E_{\rm g}^{\rm app}(\xi)$ (blue solid line). In the inset we represent the corresponding errors  $|\Delta E'_{\rm g}|=|E_{\rm g}^{\rm app}-E_{\rm g}^{\rm DMRG}|$ (blue-red crosses), and $|\Delta E'_{\rm g}|=|E_{\rm g}^{\rm JW}-E_{\rm g}^{\rm DMRG}|$ (yellow-red crosses). }
\label{scaling_dmrg}
\end{figure}

 Finally, we have checked the scaling of the groundstate energy with the dimerization parameter~\eqref{gs_lowering}. We consider a spin chain with $L=200$ sites, and use the finite-size DMRG algorithm for $m=50$ and $N_{\rm s}=2$. In this case, we vary the dimerization $\xi_i\in[0,0.5]$ for $g=0$, and study the behavior of the groundstate energy (see Fig.~\ref{scaling_dmrg}). The agreement shown in this figure, even for relatively large dimerizations,  provides sufficient evidence to support the claims of this work.

\subsection{Spin-Peierls Quantum Phase Transition}

In this  Appendix, we discuss the possibility of adjusting the parameters of the spin-phonon model, such that the magnetic and structural phase transitions occur simultaneously as a consequence of the quantum fluctuations introduced by the transverse field $g$. Let us recall that under the adiabatic  and homogeneous approximations for the $\phi^4$ model, the lowering of the groundstate energy~\eqref{gs_lowering} can be rewritten as follows
\begin{equation}
 E_{\rm g}(\xi)-E_{\rm g}(0)=-\frac{2J\theta^2\varphi(g)}{\pi}\sum_i\delta q_{ix}^2.
\end{equation} 
Since this expression is quadratic in the zigzag displacements, we must modify the parameters of the  scalar field theory~\eqref{r_app} 
  \begin{equation}
  \label{displaced_spt}
  \begin{split}
  r^x_{i}&\to \tilde{r}^x_{i}=m\omega_x^2l_z^2\left(1-\kappa_x\left(\frac{\zeta_i(3)}{2}+\frac{2J}{m\omega_z^2l_z^2}\frac{\theta^2\varphi(g)}{\pi}\right)\right),\\
    u^x_{i}&\to \tilde{u}^x_{i}=u^x_{i}.
  \end{split}
 \end{equation}
In the homogeneous approximation, whereby the coupling between the neighboring double-welled oscillators is neglected,  the SPT occurs when $ \tilde{r}^x_{i}<0, \tilde{u}^x_{i}>0$. Hence, it becomes evident that the critical point is displaced towards a smaller value of the ratio between the trapping frequencies
 \begin{equation}
 \label{scp}
 \kappa_{{\rm c},i}\to\tilde{\kappa}_{{\rm c},i}=\frac{1}{\left(\frac{\zeta_i(3)}{2}+\frac{2J}{m\omega_z^2l_z^2}\frac{\theta^2\varphi(g)}{\pi}\right)}.
 \end{equation}
Hence, the lowering of the energy due to the magnetic order compensates the energy increase due to the zigzag distortion, and the linear chain becomes unstable towards the distorted lattice for $\tilde{\kappa}_{{\rm c},i}\leq \kappa_x<\kappa_{{\rm c},i}$. 

The idea now is to tailor the parameters of the model in order to make the critical points of the magnetic~\eqref{mcp} and structural~\eqref{scp} phase transitions coincide. This occurs for
 \begin{equation}
\frac{2J}{m\omega_z^2l_z^2}=\left(\kappa_x^{-1}-\frac{\zeta_i(3)}{2}\right)\frac{\pi}{\theta^2\varphi(\tilde{g}_{\rm c})},
 \end{equation}
which is equivalent to the condition for the axial trapping frequency in the main text~(15). By substituting in Eq.~\eqref{displaced_spt}, we find that
\begin{equation}
\tilde{r}^x_{i}=\half m\omega_z^2l_z^2\zeta_i(3)\left(1-\frac{\varphi(g)}{\varphi(\tilde{g}_{\rm c})}\right).
\end{equation}
Due to the monotonically-decreasing nature of the function $\varphi(g_1)<\varphi(g_2)$, for $g_1>g_2$, (see Fig.~\ref{dqim_qpt}{\bf (b)}), we find that 
 $g>\tilde{g}_{\rm c}\Rightarrow\tilde{r}_i^x>0$, leading to a disordered linear paramagnet, whereas $g<\tilde{g}_{\rm c}\Rightarrow\tilde{r}_i^x<0$ yields the ordered zigzag antiferromagnet. Hence the spin-Peierls transition is triggered by quantum fluctuations alone (i.e. spin-Peierls quantum phase transition), as outlined in the scheme of Fig.~1 in the main text.

\subsection{Spin-Peierls Quantum Simulator in Perspective}

In order to assess the scope and utility of the proposed quantum simulator, we discuss in detail some of  the experimental and theoretical efforts to understand the  spin-Peierls instability in the context of  quantum many-body physics.

{\it Experimental perspective.--} The first  experimental evidence of the spin-Peierls instability was found in the organic molecular crystal TTF-CuS$_4$C$_4$(CF$_3$)$_4$. There, unpaired spins $s=1/2$ localized at neighboring donor ions  TTF$^+$ interact via super-exchange through the acceptor  CuS$_4$C$_4$(CF$_3$)$_4$, and give rise to an alternating antiferromagnetic Heisenberg model that depends on the lattice dimerization. The opening of the spin-Peierls gap  was inferred from the sharp isotropic decrease of the magnetic susceptibility below $T_{\rm c}\approx 12$K~\cite{sp_1}, and from an anomaly in the magnetic contribution to the  specific heat~\cite{sp_3}. Besides, the associated  dimerization of the lattice was directly inferred from X-ray measurements of new Bragg peaks~\cite{sp_2}, the intensity of which  gave  evidence for  the condensation of a soft phonon mode. Finally, the clear-cut  confirmation of the magnetic origin of the structural phase transition was  obtained by applying an additional magnetic field. In the low-field regime, neutron scattering experiments displayed a shift of the critical temperature~\cite{sp_4}. Additionally, magnetization measurements showed that a new phase transition  takes place in the high-field regime~\cite{sp_5,sp_5_bis}, and leads to an incommensurate lattice distortion. Let us finally note that  spin-Peierls transitions were also observed for other organic materials, such as TTF-AuS$_4$C$_4$(CF$_3$)$_4$ ($T_{\rm c}\approx2.1$K)~\cite{sp_6}, and MEM-(TCNQ)$_2$, which displayed both an electronic Peierls transition ($T_{\rm c}\approx$335K) and a spin-Peierls instability ($T_{\rm c}\approx$18K)~\cite{sp_7}.

The discovery of a spin-Peierls phase transition in CuGeO$_{3}$ ($T_{\rm c}\approx$14K)~\cite{sp_8}, an inorganic compound, paved the way to new experiments due to the availability of larger crystals with a  higher quality. In this case, the spins $s=1/2$ of localized electrons in the Cu$^{2+}$ ions interact via super-exchange through the O$^{2-}$ ions, and also lead to an antiferromagnetic Heisenberg model that depends on the lattice dimerization. Hence,  similar effects were observed in the experiments, such as the drop of the magnetic susceptibility~\cite{sp_8}, dependence on additional magnetic fields~\cite{sp_9},  or X-ray and neutron scattering measurements of the secondary Bragg peaks~\cite{sp_10}. In addition, this new material allowed for new possibilities,  such as the measurement of the spin-Peierls gap and the low-energy dispersion relations via inelastic neutron scattering~\cite{sp_11}, or the accurate estimation of  the modified lattice constants by neutron diffraction~\cite{sp_12}. Let us remark that no evidence of the soft-phonon condensation was found in  these experiments. Another  difference with respect to the organic compounds is the presence of magnetic frustration in the underlying Heisenberg model, which is based on the analysis of magnetic susceptibility~\cite{sp_13} and magnetostriction experiments~\cite{sp_14}. 

Let us finally note that other inorganic materials, such as the transition metal oxide TiOCl~\cite{sp_15}, have also shown clear  evidence of an underlying spin-Peierls transition.  This material, which was originally proposed to realized a resonating valence bond state, has turned out to be another example of an antiferromagnetic Heisenberg magnet that displays a spin-Peierls transition at a much higher temperature $T_{\rm c}\approx 67$K.

Having described these seminal experiments, it is now possible to understand  the interest of the proposed quantum simulator from a experimental perspective. First, we note that all of the above spin-Peierls instabilities occur at a finite temperature $T_{\rm c}>0$. To the best of our knowledge, the observation of  a spin-Peierls transition only driven by quantum fluctuations is still an open problem, which could be addressed with the proposed quantum simulator. Second, we also note that the above transitions are well described by an antiferromagnetic Heisenberg model.  Hence, another open problem is the observation of spin-Peierls phenomena for materials described by other microscopic models. Unfortunately, the magnetic interaction in the above materials is determined by  a super-exchange that cannot be experimentally controlled. In  contrast, our quantum simulator would allow for the possibility of tailoring the magnetic interactions experimentally. Although we have considered an Ising-type coupling, we emphasize that different models could also be explored by exploiting other laser-induced spin-phonon couplings. Third, an additional  property of the trapped-ion quantum simulator is the capability of performing very accurate measurements at the single-particle level. In comparison to the global measurements in the above experiments, trapped-ion technologies would allow for direct measurements of the local magnetization, or the distance dependence of two-spin correlators. Besides, the crystal structure, and thus the onset of the lattice dimerization, can be directly imaged on CCD cameras. Finally, the possibility of controlling the number of ions would allow for a very interesting transition from the few- to the many-body scenario. 

{\it Theoretical perspective.--} The first theoretical work on the spin-Peierls instability~\cite{sp_t_1} conjectured the onset of a structural phase transition in molecular crystals, which would be caused by collective magnetic interactions. This was followed by the first attempts towards a microscopic theory, both for the alternating Heisenberg~\cite{sp_t_2,sp_t_3} and  XY~\cite{sp_t_4} magnets. In the latter, the spin model can be mapped onto free spinless fermions, and thus connected with previous work on the fermionic Peierls transition (see e.g.~\cite{cdw_book}). The alternating Heisenberg model~\cite{sp_t_3} was initially treated within  mean-field theory (i.e. Hartree-Fock approximation), such that the results of the XY model could be used after a simple renormalization of the spin-spin interaction strength. In particular, this mean-field theory predicts  the sharp decrease of the magnetic susceptibility below $T_c$, and was used satisfactorily to explain the experimental results on the organic crystals described above.    The surprising accuracy of the mean-field predictions could only be   understood in the light of the Luttinger-liquid treatment~\cite{sp_t_5}, which predicted only small corrections for the measured macroscopic quantities, with the possible exception of  the experiments with strong magnetic fields~\cite{sp_5_bis}.

The challenge to theoretical theories beyond the mean-field treatment changed considerably with the discovery of the inorganic spin-Peierls sample CuGeO$_3$. First,  the underlying microscopic model must incorporate a certain amount of frustration caused by next-to-nearest neighbor interactions~\cite{sp_13}. We note that the interplay of frustration and quantum fluctuations is responsible for a complex phase diagram already for the bare spin model~\cite{j1_j2_heisenberg}. In the complete spin-phonon system, the spin-Peierls instability and the frustration constitute two different mechanisms for opening an energy gap, which lead to two different types of phase transitions (i.e. second order and Kosterlitz-Thouless, respectively). Second, the lack of experimental evidence supporting a soft-phonon mode questions the validity of the so-called adiabatic approximation for the crystal vibrations. In most of the theoretical approaches above,  the lattice dynamics is neglected by only considering a static elastic distortion. Some of the first attempts to  account for the lattice dynamics considered the opposite diabatic limit~\cite{sp_t_6}, where the phonons, which are much faster than the spin excitations in this regime, contribute to a phonon-mediated interaction. In fact, this mechanism can be also responsible for the long-range frustrating interactions found in  CuGeO$_3$. Let us note, however,  that the spin-spin coupling in this material has roughly the same order of magnitude as the phonon frequency, which seems to point out to a regime which is neither adiabatic nor diabatic. In this regime, one must resort to numerical tools, such as optimized exact diagonalization~\cite{sp_t_7}, density matrix renormalization group algorithms~\cite{sp_t_8}, or quantum Monte Carlo~\cite{sp_t_9}. These numerical results showed that the non-adiabaticity leads to important renormalization of the spin-Peierls gap and the lattice dimerization with respect to the static theories. More importantly, it can also destabilize the spin-Peierls phase for strong-enough spin-phonon couplings via a  Kosterlitz-Thouless transition. 
 
 After this description, we can now discuss  the interest of the proposed quantum simulator from a theoretical point of view. First, the trapped-ion setup is an excellent playground where to study strong deviations from mean-field theories. Second, the quantum simulator incorporates naturally the effects of magnetic frustration, since the spin interactions develop a dipolar long-range. We note that the degree of frustration, namely the ratio between nearest and next-to-nearest  couplings, could be tailored by slightly tilting the laser beams in such a way that the effective wavevector also has a component along the axis of the chain. Hence, the interplay of quantum fluctuations and frustration can de actively designed in the quantum simulator. Finally, and most notably, the degree of adiabaticity can also be controlled. We have focused on a setting where the trapping frequencies lie close to the critical point of the structural transition. In this regime,  the frequency of the  soft-phonon mode  lies below the effective spin-coupling strengths, and the adiabatic approximation holds. However, the quantum simulator could in principle also explore the diabatic regime and the intermediate one by simply changing the trap frequencies. From our point of view, this is a very important property of the quantum simulator, since analytical approaches do not exist  in the intermediate regime. Besides, numerical methods may even lead to some controversy when the dispersion of the phonon modes is taken into account~\cite{sp_t_10}. We would like to note that the efficiency of the numerical methods is likely to get degraded  when the $\phi^4$ non-linearities are considered. A priori, the proposed spin-Peierls quantum simulator  has the potential of
exceeding the power of numerical methods on classical computers, which should be the final goal of any sensible quantum simulator.
 

\end{document}